\documentclass[aps,pra,groupaddress]{revtex4-1}
\usepackage{bm}
\usepackage{graphicx}
\usepackage{amsmath}

\begin{document}
\title{Towards Quantum Integrated Information Theory}
\author{Paolo Zanardi$^{1,2}$, Michael Tomka$^{1,2}$, Lorenzo Campos Venuti$^{1,2}$ }
\affiliation{$^1$ Department of Physics and Astronomy, University of Southern California, Los Angeles, CA 90089-0484, USA \\
$^2$ Center for Quantum Information Science \& Technology, University of Southern California, Los Angeles, California 90089, USA}




\begin{abstract}
Integrated Information Theory (IIT) has emerged as one of the leading
research lines in computational neuroscience to provide a mechanistic
and ma\-the\-ma\-ti\-cal\-ly well-defined description of the neural correlates
of consciousness. Integrated Information ($\Phi$) quantifies how much the integrated
cause/effect structure of the global neural network fails to be
accounted for by any partitioned version of it.
The holistic IIT approach is in principle applicable to any
information-processing dynamical network regardless of its
interpretation in the context of consciousness.
In this paper we take the first steps towards a formulation of a
general and consistent version of IIT for interacting networks of
quantum systems. 
A variety of different phases, from the dis-integrated ($\Phi=0$) to
the holistic one (extensive $\log\Phi$), can be identified and their
cross-overs studied.
\end{abstract}
\maketitle

\section{Introduction}
Over the last decade Integrated Information Theory (IIT), developed by
G.~Tononi and collaborators, has emerged as one of the leading
research lines in computational neuroscience. 
IIT aims  at providing a mechanistic and mathematically well-defined
description of the neural correlates of consciousness~\cite{IIT-0, IIT-1,IIT-2,IIT-3}. 

The idea is to quantify the amount of cause/effect power in the neural
network that is holistic in the sense that goes beyond and above the
sum of its parts. This is done in a bottom-up approach by quantifying how arbitrary
parts of the network (``mechanisms''), in a given state, influence the
future and constrain the past of other arbitrary parts (``purviews''), in
a way that is irreducible to the separate (and independent) actions
of parts of the mechanism over parts of the purview. 
Iterated at the global network level this process gives rise to a
so-called ``conceptual structure", comprising a family of mechanisms and
purviews, where the latter represent the integrated core causes/effects of the former~\cite{IIT-1}.
A measure of the distance between this conceptual structure with the closest one obtainable from a suitably partitioned network quantifies  how much of  the cause/effect structure of dynamical  newtwork fails to be reducible to the sum of its parts. This minimal distance is, by definition,  the Integrated Information (denoted by $\Phi$) of the network.

In IIT it is then boldly postulated that the larger $\Phi$, the higher
is the degree of consciousness of the network in the given state.
The irreducibility of the causal information-processing structure of
the network measured by $\Phi$ is independent of the specific
``wetware" implementing the brain circuitry.
It follows that the IIT approach to consciousness seems to 
lead to a, rather controversial~\cite{scott},  panpsychist view of the world~\cite{IIT-3}.

Besides, and irrespective of, the applications to
consciousness, the IIT approach is in principle applicable to
any information-processing  network.
For example, applications of IIT to Elementary Cellular Automata and
Adapting Animats have been discussed~\cite{Larissa}.
Moreover, potential extensions of IIT to more general systems,
including quantum ones, have been proposed in~\cite{Max-1,Max-2}
by M.~Tegmark (see also~\cite{Ranchin2015}).

{{In this paper  we shall make an attempt to formulate a general and
consistent version of IIT for interacting networks of finite-dimensional and non-relativistic quantum systems.}}
Our approach is going to be a quantum information-theoretic one: neural networks are being replaced by networks of qudits, probability
distributions by non-commutative density matrices, and markov processes by trace preserving completely positive maps. The irreducible cause/effect structure of the global network is encoded by a so-called conceptual structure operator. The minimal distance of the latter from those obtained by factorized versions of the network, defines the quantum Integrated Information $\Phi$. 
We would like to strongly emphasize from the very beginning that:

 {\bf{i)}} Our goal is {\em{not}} to account for potential quantum features of consciousness. We aim at understanding the role that, a suitably designed notion of,  information integration may play in {\em{a)}} quantum information processing in {\em{sensu lato}}, and {\em{b)}}  in a novel categorization of the different phases of quantum matter.

{\bf{ii)}} The quantum extension of IIT  (QIIT) that we are going to discuss is not unique. In fact, exploring new avenues toward QIIT  is one of the main goals for further investigations.

Here we deem necessary a word of warning: in the following we will often borrow jargon from  classical IIT e.g., {\em{mechanism, repertories, purviews, concepts, conceptual structures,...}}, these are technical terms  (precisely defined in the paper) which may  not be necessarily familiar to quantum information experts  and should {\em{not}} be confused with the ordinary language usage of the same terms.

\section{Setting the stage}
Let $\Lambda$ be a set of cardinality $|\Lambda|<\infty$.
For each $j\in\Lambda$ there is an associated $d-$dimensional quantum
system with Hilbert space $h_j \cong {\mathbf{C}}^d$.
Adopting the IIT jargon we will refer to subsets $M$ of $\Lambda$ as to {\em{mechanisms}}.
Usually $\Lambda$ will be equipped by a distance function $d;$ in this case we define the distance between mechanisms $M$ and $P$ by:
${\mathrm{dist}}(M, P):={\mathrm{min}}_{x\in M,\, y\in P} d(x,y).$
Given any $\Omega\subset\Lambda$ we define 
${\cal H}_\Omega=\otimes_{j\in\Omega} h_j$, with dimension
$d^{|\Omega|}$.
We will denote by $L({\cal H}_\Lambda)$  (${\cal S}({\cal H}_\Lambda)$) the associated
operator-algebra (state-space).
One has that
${\cal H}_\Lambda\cong {\cal H}_\Omega \otimes {\cal H}_{\Omega^\prime}$,
where $\Omega^\prime$ denotes the complement of $\Omega$ (in
$\Lambda$).
The network dynamics will be described by a trace preserving {\em{unital}} CP-map 
${\cal U}\colon L({\cal H}_\Lambda) \rightarrow  L({\cal H}_\Lambda)$
with ${\cal U}({\mathbf{1}})={\mathbf{1}}$.
This map has to be thought of as the one-step evolution of a discrete
time process. If $\cal U$ is a CP-map its dual ${\cal U}^*$ is defined by: $\langle X, {\cal U}(Y)\rangle= \langle {\cal U}^*(X), (Y)\rangle,\,\forall X, Y\in L({\cal H}_\Lambda).$
Given $\Omega\subset\Lambda$ we define the {\em{noising}} CP-map
${\cal N}_\Omega$ by
${\cal N}_\Omega\colon L({\cal H}_\Lambda)\rightarrow L({\cal H}_\Lambda)\colon X\mapsto ({\mathrm{Tr}}_{\Omega} X) \otimes
\frac{ {\mathbf{1}}_{\Omega}} {d^{|\Omega|}} \label{noising}$. 

%
\begin{figure}[h!]
\includegraphics[scale=0.25]{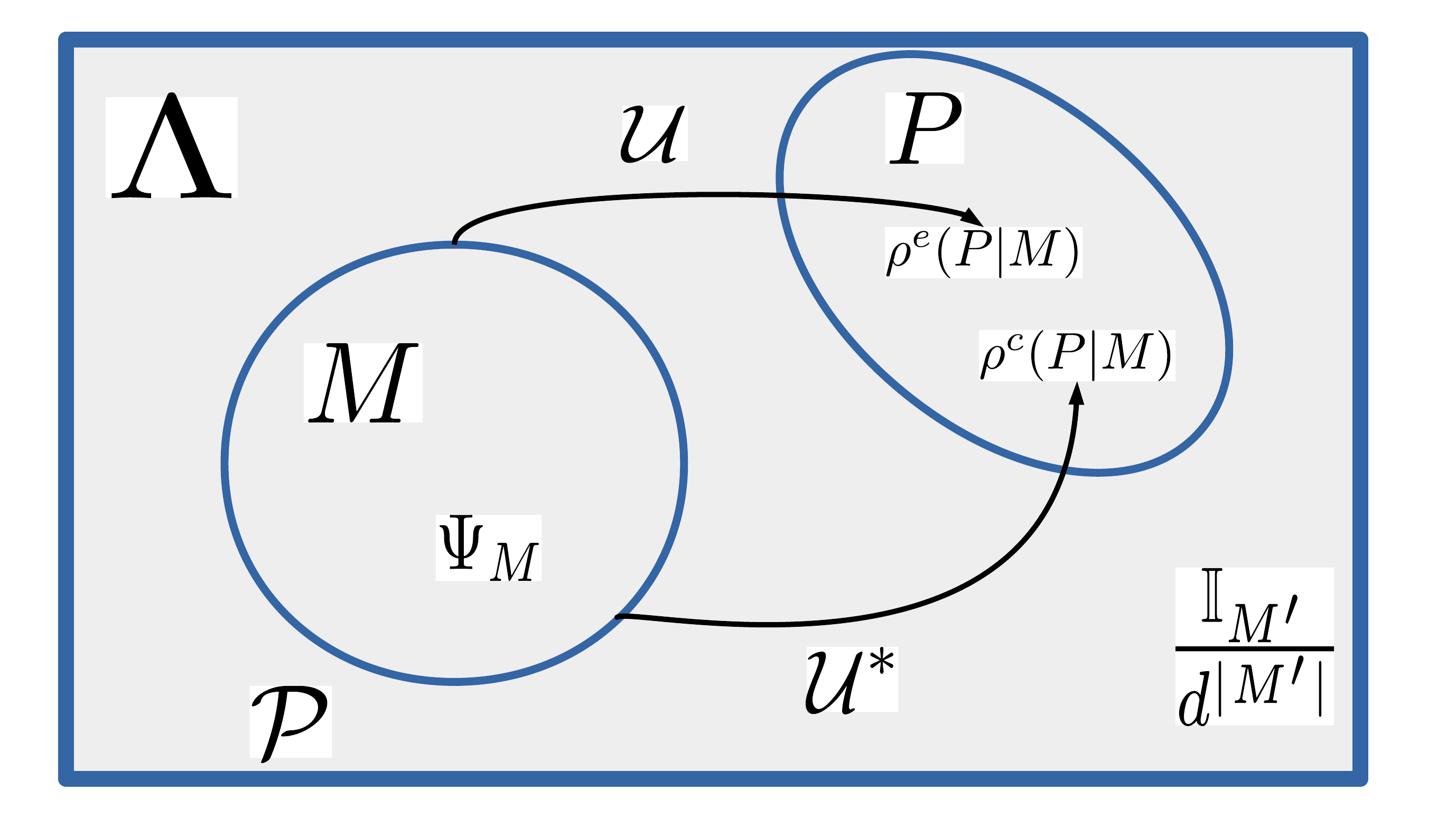}
\caption{{\small{
 The {\em{mechanism}} $M$ in state $\Psi_M$ conditions (constrains) the future (past) of the {\em{purview}} $P$ by means of the action of $\cal U$ (${\cal U}^*$).
The complement $M^\prime$ of $M$ is ``noised" and set into the maximally mixed state  $\frac{  {\mathbf{1}}_{M^\prime}  }{ d^{|M^\prime|}}.$}}
}
\label{fig:mech-pur-fig}
\end{figure}
The first step in classical IIT is to is to consider pairs $M, P\subset \Lambda$ (where the  mechanism $P$ is referred to as the {\em{purview}} of $M$)
and to quantify how the state i.e., a probability distribution, of $M$ (with $M^\prime$ being in a maximally random state) at time $t$ conditions (constraints) the state of $P$ (at time $t-1$). The quantification is obtained 
measuring the distance between the conditioned states (referred to as the effect and cause {\em{repertoires}} of $P$) and the un-conditioned one. 
It follows that  the first step toward a QIIT of is to define a quantum version of the cause/effect
repertoires of classical IIT~\cite{IIT-2,IIT-3}.
Let us now  motivate our choice for the quantum counterparts.

{\em {Effects.--}}
We will denote by $p$ ($p^\prime$) the degrees of freedom (DOFs)
associated with the purview $P$ at time $t+1$ (its complement
$P^\prime$) and by $m$ ($m^\prime$) the DOFs associated to the
mechanism $M$ at time $t$ (its complement $M^\prime$).
On purely classical probabilistic grounds one can write
\begin{eqnarray}
{\mathrm{Pr}}(p|m)&=&\frac{ {\mathrm{Pr}}(q,m) }{{\mathrm{Pr}}(m)} =\sum_{p^\prime,m^\prime}\frac{ {\mathrm{Pr}}(q,q^\prime, m, m^\prime) }{{\mathrm{Pr}}(m)} \nonumber \\
&=& \sum_{p^\prime,m^\prime}\frac{ {\mathrm{Pr}}(q,q^\prime, m, m^\prime) }{{\mathrm{Pr}}(m) {\mathrm{Pr}}(m^\prime)  }{\mathrm{Pr}}(m^\prime)=
\sum_{p^\prime,m^\prime}{ {\mathrm{Pr}}(q,q^\prime| m, m^\prime) }{\mathrm{Pr}}(m^\prime).
\end{eqnarray}
Here we have assumed that the prior of $m$ and $m^\prime$ factorizes,
i.e.,
${\mathrm{Pr}}( m, m^\prime)=  {\mathrm{Pr}}( m) {\mathrm{Pr}}( m^\prime)$. 
Now quantum mechanics enters in defining the transition probability
${\mathrm{Pr}}(q,q^\prime| m, m^\prime)= \langle p, p^\prime|{\cal U}(|m,m^\prime\rangle\langle m,m^\prime |)|p, p^\prime\rangle$,
where ${\cal U}$ is the unital CP-map describing the (one-step)
dynamics of the network.
Inserting this in the equation above one gets
\begin{eqnarray}
{\mathrm{Pr}}(p|m)&=& \sum_{p^\prime} \langle p|\langle p^\prime| {\cal U}(|m\rangle\langle m|\otimes \sum_{m^\prime}    {\mathrm{Pr}}( m^\prime) |m^\prime\rangle\langle m^\prime|  )|p^\prime\rangle |p\rangle \nonumber \\
& =&\langle p| {\mathrm{Tr}}_{P^\prime} {\cal U}(|m\rangle\langle m|\otimes \frac{  {\mathbf{1}}_{M^\prime}  }{ d^{|M^\prime|}   })|p\rangle.
\label{effect-rep}
\end{eqnarray}
Here we have assumed that the prior for $m^\prime$ is the uniform
unconstrained one, i.e., ${\mathrm{Pr}}(m)=d^{-|M^\prime|}$.
The last equation shows that the probability for the purview being in
the state $p$ at time $t+1$
(conditioned on the mechanism being in the state $m$ at time $t$)
is the diagonal element $|p\rangle$ of the reduced density matrix
$\rho_{\cal U}(P|M):=  {\mathrm{Tr}}_{P^\prime} {\cal U}(|m\rangle\langle m|\otimes \frac{  {\mathbf{1}}_{M^\prime}  }{ d^{|M^\prime|}   })$.

{\em {Causes.--}} We now consider cause repertoires and denote by $p$
($p^\prime$) the degrees of freedom (DOFs) associated with the purview
$P$ at time $t-1$ (its complement $P^\prime$) and by $m$ ($m^\prime$)
the DOFs associated to the mechanism $M$ at time $t$ (its complement
$M^\prime$).
Using Bayes rule and Eq.~(\ref{effect-rep}) (with $p$ and $m$
interchanged) one can write
\begin{eqnarray}
{\mathrm{Pr}}(p|m)&=&\frac{ {\mathrm{Pr}}(m|p)  {\mathrm{Pr}}(p)      }{{\mathrm{Pr}}(m)}=  \langle m| {\mathrm{Tr}}_{M^\prime} {\cal U}(|p\rangle\langle p|\otimes \frac{  {\mathbf{1}}_{P^\prime}  }{ d^{|P^\prime|}   })|m\rangle
\frac{{\mathrm{Pr}}(p) }{{\mathrm{Pr}}(m)} \nonumber \\
&=&{\mathrm{Tr}}\left[ (|m\rangle\langle m|  \otimes \frac{  {\mathbf{1}}_{M^\prime}  }{ d^{|M^\prime|}   } )\, {\cal U}(|p\rangle\langle p|\otimes
   {\mathbf{1}}_{P^\prime})               \right].
\end{eqnarray}
Here we used
$ \frac{{\mathrm{Pr}}(p) }{   d^{|P^\prime|}   {\mathrm{Pr}}(m) }=d^{
  |M|-|P|-|P^\prime|  }=d^{-(|\Lambda|-|M|)}=d^{-|M^\prime|}$.
Using the Hilbert-Schmidt dual
 and the properties of reduced density
matrices, the equation above becomes
\begin{eqnarray}
{\mathrm{Pr}}(p|m)&=&{\mathrm{Tr}}\left[ {\cal U}^*(|m\rangle\langle m|  \otimes \frac{  {\mathbf{1}}_{M^\prime}  }{ d^{|M^\prime|}   } )\,(|p\rangle\langle p|\otimes
   {\mathbf{1}}_{P^\prime}) \right] \nonumber \\
&=&\langle p| {\mathrm{Tr}}_{P^\prime} {\cal U}^* (|m\rangle\langle m|  \otimes \frac{  {\mathbf{1}}_{M^\prime}  }{ d^{|M^\prime|}   } ) |p\rangle.
\end{eqnarray}
Again, the last equation shows that the probability for the purview
being in the state $p$ at time $t-1$ 
(conditioned on the mechanism being in the state $m$ at time $t$)
is the diagonal element $|p\rangle$ of 
$\rho_{{\cal U}^*}(P|M)$.
%
\section{QIIT}
The considerations above naturally lead  to a definition for cause/effect repertoires where  we consider the full quantum density matrix as
opposed to just its diagonal entries. 
We would like to stress that, given the essential role of   entanglement in quantum theory, in our approach we drop out the assumption of {{conditional independence}} of the repertoires and
  the associated need of {{virtualization}}~\cite{IIT-2}.
Also, notice that in this paper we restrict ourselves to the unital case, in order to have the unconditioned repertoires equal to the maximally mixed state (see below).
This is a simplifying technical assumption, not a key requirement. 
\vskip 0.2truecm
{\bf{Definition 1a: cause/effect Repertoires:}}
Given the unital $\cal U$, the state
$\Psi_\Lambda\in {\cal S}({\cal H}_\Lambda),$
and $M, P \subset \Lambda$, we define the {{effect (e) and cause (c) repertoire}}
of $M$ over the {{purview}} $P$, by
\begin{equation}
\rho^{(x)}(P|M):= {\mathrm{Tr}}_{P^\prime}\, {\cal U}^{(x)}\circ{\cal N}_{M^\prime} (\Psi_\Lambda)=   
 \rho^{(x)}(P|M):= {\mathrm{Tr}}_{P^\prime}\, {\cal U}^{(x)}\left(\Psi_M\otimes \otimes \frac{  {\mathbf{1}}_{M^\prime}  }{ d^{|M^\prime|}}\right),\quad (x=e,c)
\label{repertoire}
\end{equation}
where, $\Psi_N=  {\mathrm{Tr}}_{M^\prime}\Psi_\Lambda,$ ${\cal U}^{(e)}= {\cal U}$ and ${\cal U}^{(c)}={\cal U}^*$ (Hilbert-Schmidt {{dual}} of ${\cal U})$.
\vskip 0.2truecm
The set of density matrices $\rho^{(e)}(P|M)$ ($\rho^{(c)}(P|M)$) encode how the dynamics constrains the future (past) of $P$, given that
the system is initialized in $\Psi_{M}$ and noised over $M'$ (see Fig.~(\ref{fig:mech-pur-fig})).
From a qualitative physical point of view one might think in the following way: $\Lambda$ supports an extended quantum medium
that is everywhere at infinite temperature but over the region $M$ where it has been locally ``cooled off" to some (possibly pure) quantum state $\Psi_M.$
The system is then evolved forward (backward) in time by the map $\cal U$ (${\cal U}^*$).  The quantities (\ref{cei}) quantifies the distinguishability of the states obtained in  this
way from the infinite temperature one if only measurements local to the region $P$ are allowed. 

The next step is to define the cause/effect information by the information-theoretic distance between the conditioned and the un-conditioned repertoire $\rho^{(x)}(P|\emptyset)={\cal U}^{(x)}(\frac{{\mathbf{1}}_{\Lambda}} {d^{|\Lambda|}})=\frac{ {\mathbf{1}}_{P}} {d^{|P|}},\,(x=e,c).$
In classical IIT the distance between repertoires 
is usually taken to be the Wasserstein distance~\cite{IIT-2}.
In this paper, in view of its salient quantum-information theoretic
properties and simplicity, we will adopt the trace distance between
density matrices $\rho$ and $\sigma$ as a measure of
statistical distinguishability, i.e., 
$D(\rho,\sigma):=\frac{1}{2}\|\rho-\sigma\|_1\in[0,1]$.
\vskip 0.2truecm
{\bf{Definition 1b: cause/effect Information:}}
The cause/effect {{information}} of $M$ over $P$ is given by 
\begin{equation}
xi(P|M):= D(\rho^{(x)}(P|M), \frac{ {\mathbf{1}}_{P} } {d^{|P|}}),\quad (x=e,c).
\label{cei}
\end{equation}
\vskip 0.2truecm
{\bf{A:  Repertories for the Swap operation}}
For the sake of illustration we will us the case $d=2=|\Lambda|$ with ${\cal U}(X)=S XS$ where $S\colon {\mathbf{C}}^2\otimes {\mathbf{C}}^2\rightarrow {\mathbf{C}}^2\otimes {\mathbf{C}}^2/\phi\otimes \psi\mapsto \psi\otimes\phi,$  is a {\em{swap}} operation. The initial state is taken in the factorized form $\Psi_\Lambda=\Psi\otimes \Psi$ where $\Psi$ is a pure density matrix over ${\mathbf{C}}^2.$ One can easily check that the non-trivial repertoires (notice that ${\cal U}={\cal U}^*$) are given by
$\rho(1|1)=\frac{ {\mathbf{1}} }{2},\,\rho(2|1)=\Psi,\, \rho(\Lambda|1)=\frac{ {\mathbf{1}} }{2}\otimes\Psi,\,
\rho(2|2)=\frac{ {\mathbf{1}} }{2},\,\rho(1|2)=\Psi,\, \rho(\Lambda|2)=\Psi\otimes \frac{ {\mathbf{1}} }{2},\, \rho(1|\Lambda)=\Psi,\, \rho(2|\Lambda)=\Psi,$ and $\rho(\Lambda|\Lambda)=\Psi^{\otimes\,2}.$ From this it follows $xi(1|1)=xi(2|2)=0,\, xi(2|1)=xi(1|2)=xi(1|\Lambda)=xi(2|\Lambda)=\frac{1}{2},\,xi(\Lambda|\Lambda)=\frac{3}{4}.$
\vskip 0.2truecm
At the technical level the following remarks  are now useful:

{\bf{1)}} 
Since pure states have the maximum distance from the maximally mixed
state and by distance monotonicity under partial traces 
$xi(P|M)\le {\mathrm{min}} \{ 1-d^{-|P|},\,1- d^{-|M|}\}.$


{\bf{2)}} For  unitary $\cal U$'s generated by a local Hamiltonian $H_\Lambda=\sum_{X\subset\Lambda,\,|X|=O(1)} H_X,$ 
the functions $xi(P|M),\,(x=e,c)$ fulfill a Lieb-Robinson type 
inequality~\cite{LR} 
$$xi(P|M)\le c \exp\left( -a (\mathrm{dist}(P,M)- v|t|)\right),\, (x=c,e).$$ 
Here $a, c>0$ are constants depending on $d_M, |M|, |P|,$
and $\|O_P\|.$ Moreover, $v>0$ is the Lieb-Robinson velocity which depends on $H_\Lambda$ (see Appendix for a proof).  

{\bf{3)}} The {\em{average cause/effect information}} of a map $\cal U$ is defined by the uniform average of $xi(P|M)$ over all mechanisms/purviews
$XI({\cal U}):= \frac{1}{2^{2|\Lambda|}}\sum_{P,M\subset\Lambda} xi(P|M)=:\langle xi(P|M) \rangle_{P,M},\, (x, X=c,e)$

{\bf{4)}} Using the inequality $\|\rho-\sigma\|_1^2\le 2 S(\rho||\sigma)$ one finds 
$xi(P|M)\le \frac{1}{\sqrt{2}}\sqrt{S^{max}_P -S(\rho^{(x)}(P|M))},\,(x=e,c)$ 
where $S^{max}_P :=\log d^{|P|},$  here $S$ denotes the von-Neumann entropy. Introducing the $2$-Renyi entropy i.e., $S_2(\rho)=-\log {\mathrm{Tr}}(\rho^2)\le S(\rho) $ one gets
\begin{equation}
 xi(P|M)\le\sqrt{\frac{1}{2}\log\left(d^{|P|}\,\|\rho^{(x)}(P|M)\|_2^2\right)},\quad(x=c,e).
\label{upper-bound-cei}
\end{equation}
This inequality is useful as the purity $\|\rho^{(x)}(P|M)\|_2^2$ is technically easier to handle than the trace-distance.  

Let us illustrate this fact in two ways: the first shows that the conditional repertoires purities have a simple expression in terms of standard multi-point spin correlators;
the second shows how, using Eq.~(\ref{upper-bound-cei}), one can gain an insight on the behavior of cause-effect power for typical (Haar) random unitaries.

\vskip 0.2truecm
{\bf{5)}}
We focus on effect repertoires as everything in the following holds for cause ones by replacing ${\cal U}$ with ${\cal U}^*.$ Using the notation $\{\sigma^{(j)}_\alpha\}_{\alpha=0}^3= \{{\mathbf{1}},\sigma_x^{(j)},  \sigma_y^{(j)},\sigma_z^{(j)}\}$ for the $j$- th spin (tensorized with the identity 
over $\Lambda-\{j\}$) one finds \cite{unp} $\Psi_M=\otimes_{j\in M} |\psi_j\rangle\langle \psi_j|= \prod_{j\in M}\frac{1}{2}({\mathbf{1}} +\lambda^{(j)}\cdot \sigma^{(j)})=2^{-|M|}\sum_{\beta\in{\mathbf{Z}}_4^{|M|}}
\prod_{j\in M} \lambda^{(j)}_{\beta_j} \sigma_{\beta_j}^{(j)}$ and 
\begin{eqnarray}
\|\rho^{(e)}(P|M)\|_2^2=
\frac{1}{2^{|P|}} \sum_{\alpha\in{\mathbf{Z}}_4^{|P|}}|\sum_{\beta\in{\mathbf{Z}}_4^{|M|}} G(P|M)_{\alpha, \beta} \lambda_\beta|^2,
\end{eqnarray}
where $\lambda_\alpha= \prod_{j\in M} \lambda^{(j)}_{\alpha_j}$ (similarly for $\lambda_\gamma$) and 
$$G(P|M)_{\alpha, \beta}:= \frac{1}{2^{|\Lambda|}} {\mathrm{Tr}}\left[ \prod_{j\in P}\sigma^{(j)}_{\alpha_j} \,{\cal U}\left( \prod_{j\in M}  \sigma_{\beta_j}^{(j)}\right)\right],\,
(\alpha\in{\mathbf{Z}}_4^{|P|}, \beta\in {\mathbf{Z}}_4^{|M|})$$
is a $(|P|+|M|)$-point (infinite temperature) spin-spin correlator for the CP-map ${\cal U}.$ Similiar expressions hold for the cause repertoires.
%

In the special case $|M|=|P|=1$ (i.e., both mechanism and its purview consist of  single qubit, say the $i$-th and the $j$-th respectively) ) one has a further simplification. Indeed
in this case $G(j|i)_{0,0}=1$ and $G(j|i)_{0,\beta}=G(j|i)_{\beta, 0}=0,\,(\beta=1,2,3)$ from which it follows  
$\|\rho^{(e)}(j|i)\|_2^2=\frac{1}{2} (1 +\|G(j|i) \lambda^{(i)}\|^2)$ where $G(j|i)_{\alpha,\beta}=2^{-|\Lambda|}\, {\mathrm{Tr}}(\sigma^{(j)}_\alpha{\cal U}(\sigma^{(i)}_\alpha))$
is $3\times 3$ two-point spin correlator, $\lambda^{(i)}$ is the Bloch vector of the mechanism state and $\|\bullet\|$ denotes the standard euclidean norm.
Moreover the Bloch vector of $\rho^{(e)}(j|i)$ is nothing but $G(j|i) \lambda^{(i)}$ i.e.,    $ \rho^{(e)}(j|i)=\frac{1}{2}[{\mathbf{1}}+ (G(j|i) \lambda^{(i})\cdot\sigma^{(j)}]. $

{\bf{6)}} For unitary evolutions $U$ and  pure and factorized $\Psi_\Lambda,$ one can explicitly (Haar) average over $U$'s ~\cite{unp}
\begin{equation}
{\mathbf{E}}_U\left[
\|\rho_U^{(x)}(P|M)\|_2^2\right]=\frac{1}{2}
\sum_{\alpha=\pm 1}\left(  \frac{d^{|\Lambda|}+\alpha d^{|M|} }{d^{|\Lambda|}+\alpha}\right)\left( \frac{1}{d^{|P|}} +\alpha \frac{1}{d^{|P^\prime|}} \right),\quad(x=e,c)
\label{purity-average}
\end{equation}
This result is the same for cause and effects repertoires (invariance of the Haar measure under $U\mapsto U^\dagger$) and its state independent.
If $|P|=O(1)$ and (i.e., the purview  not a finite fraction of $|\Lambda|$) then from (\ref{purity-average})  it follows  that 
${\mathbf{E}}_U\left[{d^{|P|}\|\rho_U^{(x)}(P|M)\|_2^2}\right]=1+O(e^{-|\Lambda|})$ which in turn, using (\ref{upper-bound-cei}) and concavity,  implies 
${\mathbf{E}}_U\left[xi_U(P|M)\right]=O(e^{-|\Lambda|/2}).$
This bound holds true for {\em{any}} mechanism $M.$ For $|M|=O(\Lambda)$ the physical interpretation is that a  typical (Haar) random $U$ will map the initial network state onto a nearly maximally entangled one which locally, for $|P|=O(1),$ will look almost indistinguishable from the maximally mixed state  i.e., the unconditional one.
This remark may seem to  suggest that quantum {\em{entanglement}} plays a sort of ``negative" role in the type of QIIT we are here trying to develop (see more about this issue later on).
\vskip 0.2truecm
{\bf{Examples.}} The following examples show that the  type of causal power defined by Eq.~(\ref{cei}) has counter-intuitive aspects and should, therefore, handled with care.
When ${\cal U}={\cal U}^*={\mathbf{1}}$  one has that $\rho^{(x)}(P|M)=\Psi_{P\cap M}\otimes \frac{ {\mathbf{1}}_{P\cap M^\prime}} {d^{|P\cap M^\prime|}};$
now if  $\Psi_\Lambda=\otimes_{i\in\Lambda} |\psi_i\rangle\langle \psi_i|$ one finds $
xi(P|M)=1-d^{-|P\cap M|},\,(x=c,e)$. Moreover the XI can be computed using the fact that $|P\cap M|=\sum_{i\in\Lambda} x_M(i) x_P(i)$ 
\begin{multline}
XI({{\mathbf{1}}})= \frac{1}{2^{2|\Lambda|}}\sum_{x_P,x_M\in\{0,1\}^{|\Lambda|} } (1-d^{-\langle x_M, x_P\rangle})= \\
1-\frac{1}{2^{2|\Lambda|}}\sum_{x_P,x_M\in\{0,1\}^{|\Lambda|} } \prod_{i\in\Lambda}d^{-x_M(i) x_P(i)}
=1-\left(\frac{3d+1}{4d}\right)^{|\Lambda|}. 
\label{XI-Id}
\end{multline}
Here $x_{M,P}\in \{0,1\}^{|\Lambda|}$ are  bit-strings of length $|\Lambda|$ which parametrize the sets $M$ and $P.$ Notice that the {\em{same}} result holds for {\em{any}} totally  factorized unitary $U=\otimes_{i\in\Lambda} U_i.$
For one qubit one has $XI({{\mathbf{1}}})_{d=2,|\Lambda|=1}=1-7/8=1/8;$ whereas for two qubits $XI({{\mathbf{1}}})_{d=2,|\Lambda|=2}=1-49/64=15/64.$
The latter result is identical to the one for $U$ being the swap between the two qubits (direct computation) showing that the XI of identity can be equal to the one of non trivial
(and integrated) transformation. Moreover, for two qubits,  and  $U={\mathrm{cNOT}}$ with $\Psi=|1\rangle\langle 1|^{\otimes\,2}$ one finds (direct computation) $XI({\mathrm{cNOT}})=11/64$ showing that a non-trivial interaction can have {\em{less}} total cause-effect power that identity (i.e., doing nothing).
%
%
\vskip 0.2truecm
The next definition captures quantitatively the notion of {\em{irreducibility}}  of c/e repertories,
namely how far the conditional repertoires are from those obtainable from disjoint parts of $M$  {\em{independently}} conditioning disjoint parts of $P$.
The idea of IIT is that only irreducible actions are ``real" and exists {\em{per se}} \cite{IIT-1}. 
\vskip 0.2truecm
{\bf{Definition 2: Integrated information for mechanisms-}}
Given the mechanism $M$ and the purview $P$ we consider all possible
bi-partitions of them $\{ M_1, M_2\}$ and
$\{P_1, P_2\}$, where
$X_1 \cap X_2 = \emptyset, \, X_1 \cup X_2 = X \, (X = M, P)$.
We define the (cause/effect) {{integrated  information}} (ii) of $M$
over $P$ by
\begin{equation}
\varphi^{(x)}(P|M) = \min_{(P_i, M_i)} D[\rho^{(x)}(P|M),
\rho^{(x)}(P_1|M_1)\otimes\rho^{(x)}(P_2|M_2)] \in  [0,1], \, (x=e,c)
\label{varphi}
\end{equation}
\vskip 0.2truecm
In this definition the minimum is taken over all the $2^{|P|+|M|-1}-1$
possible pairings $(P_i, M_i)\,(i=1,2)$ different from the trivial one
$(\emptyset, \emptyset),\,(P,M)$, which would make any repertoire factorizable. 
Notice that, since the $\rho^{(x)}(P_i|M_i)$'s ($i=1,2$) are
{\em{not}} the reduced density matrices of $\rho^{(x)}(P|M),$ the
factorizability of the latter is a necessary, but {\em{not}}
sufficient condition for the vanishing of $\varphi^{(x)}(P|M)$.
If $(P_1, P_2)$ is the partition of $P$, which
achieves the minimum, then quantum  entanglement of $\rho^{(x)}(P|M)$,
measured by its distance from the set of separable states over
${\cal H}_{P_1}\otimes {\cal H}_{P_2}$, provides a lower-bound to
$\varphi^{(x)}(P|M)$.  
Moreover, cause/effect information gives an upper bound to the
integrated information 
 (note that $\rho^{(x)}(\emptyset|M)=1,\,\forall M$ by normalization)
\begin{equation}
\varphi^{x}(P|M) \le D(\rho^{(x)}(P|M),  \rho^{(x)}(\emptyset|M)\otimes \rho^{(x)}(P|\emptyset))  =xi(P|M), \qquad(x=e,c).
\label{upper-bound-ii}
\end{equation}
The bound is saturated in $|M|=|P|=1.$ In particular, Eq.~(\ref{upper-bound-ii}) and remark {\bf{2)}} above
imply that integrated information obeys a Lieb-Robinson type of bound
for $\cal U$'s generated by local-Hamiltonians. This shows that $\varphi$
obeys locality in the usual sense allowed in non-relativistic quantum theory \cite{LR}.
Furthermore, Eq.~(\ref{upper-bound-ii}) along with the bounds for cause/effect information in {\bf{4)}} above show
that for finite purviews and typical (Haar) random unitaries integrated information is exponentially small in the network size.
\vskip 0.2truecm
{\bf{B: $\varphi$ for the Swap}}
From Eq.~(\ref{upper-bound-ii}) one sees that $\varphi(1|1)=\varphi(2|2)=0.$ Moreover from: $\rho(1|\Lambda)=\rho(1|2)\otimes \rho(\emptyset|1),\,
\rho(2|\Lambda)=\rho(2|1)\otimes \rho(\emptyset|2),\,\rho(\Lambda|1)=\rho(1|\emptyset)\otimes \rho(2|1),\,  \rho(\Lambda|2)=\rho(1|2)\otimes \rho(2|\emptyset)$ and $\rho(\Lambda|\Lambda)= \rho(1|2)\otimes\rho(2|1),\Rightarrow \varphi(\Lambda|1)=\varphi(\Lambda|2)=\varphi(1|\Lambda)=\varphi(2|\Lambda)=\varphi(\Lambda|\Lambda)=0.$ Finally, $\varphi(2|1)=\varphi(1|2)=\frac{1}{2}\|\Psi - \frac{ {\mathbf{1}}}{2}\|_1=\frac{1}{2}.$
\vskip 0.2truecm
Using Eq.~(\ref{varphi}) one can now, for each mechanism $M\subset \Lambda,$ identify two purviews over which $M$ has maximal irreducible causal power.
\vskip 0.2truecm
{\bf{Definition 3: Core causes, effects.--}} 
The purview $P^{(e/c)}_*$ is a {{core effect/cause}} of $M$, if 
$P^{(e/c)}_*={\mathrm{arg\,max}}_{P} \,\varphi^{(e/c)}(P|M)$.
The corresponding value of $\varphi$ will be denoted by 
$\varphi^{(x)}(M):=\max_P \varphi^{(x)}(P|M) =\varphi(P^{(x)}_*|M),\,(x=e,c)$.
The associated (global) repertoires are given by
$\rho^{(x)}(M):=\rho^{(x)}(P^{(x)}_*|M)\otimes\frac{ {\mathbf{1}}_{Q^{(x)}} }{ d^{|{Q^{(x)}| }} }
$,
where
$Q^{(x)}:=(P^{(x)}_*)^\prime$
is the complement of the core effect/cause of $M.$
The {{integrated cause/effect information}} of $M$ is given by
$\varphi(M)=\min\{\varphi^{e}(M),\, \varphi^{c}(M)\}$. 
 \vskip 0.2truecm
If $\varphi(M)=0$, then either $\varphi^{(e)}(P|M)=0,\forall P$ or
$\varphi^{(c)}(P|M)=0,\forall P$.
In the first (second) case, the mechanism $M$ fails to constrain the
future (past) on any purview $P$ in an integrated fashion.
Either way, such a mechanism is not regarded as an integrated part of
the network and it is dropped out of the picture.

The irreducible causal structure of the network has been so far described at the level of mechanisms. The next definition is instrumental in uplifting
 the construction to  the {\em{global}} network level.
\vskip 0.2truecm
{\bf{Definition 4: Conceptual Structure operators.--  }}
For any mechanism $M\subset\Lambda$ the triple
$(\rho^{c}(M), \rho^{e}(M),\varphi(M))$ with $\varphi(M)>0$ is called
a {{concept}}.
The totality of concepts forms a {\em{conceptual structure (CS)}} \cite{IIT-2}. 
Formally one can encode a CS on a positive semi-definite operator over
$({\mathbf{C}}^2)^{\otimes\,|\Lambda|}\otimes {\mathbf{C}}^2\otimes{\cal H}_\Lambda$,
given by
\begin{equation}
{C}({\cal U}):=\frac{1}{2} \,\sum_{M, \alpha}  \varphi_{\cal U}(M)\, |M \alpha \rangle\langle \alpha M|\otimes \rho_{\cal U}^\alpha(M),
\label{CS}
\end{equation}
where $M\subset \Lambda,\,\alpha=e,c,$ and we have made explicit the
${\cal U}$-dependence (but kept implicit the $\Psi_\Lambda$ one).
\vskip 0.2truecm
A CS can be also be seen a ``constellation" of triples
$ \{ (\rho^{c}(M), \rho^{e}(M),\varphi(M) )\,/\, M\subset\Lambda, \varphi(M)>0\}\subset {\cal S}({\cal H}_\Lambda)\times  {\cal S}({\cal H}_\Lambda) \times[0, 1]$. 
The latter compact set may be referred to as the quantum ``Qualia Space"~\cite{IIT-2}. 

Given two CS's, $C_1$ and $C_2$ associated to ${\cal U}_1$ and
${\cal U}_2$, respectively, we define the distance between them as the
(trace-norm) distance bewteen the associated CS operators
$D( C_1, C_2)=\frac{1}{2}\| C_1 -C_2\|_1$. More explicitly,
\begin{equation}
D( C_1, C_2)
=\frac{1}{4}\sum_{M,\alpha} \|\varphi_1(M) \rho^{(\alpha)}_1(M)- \varphi_2(M) \rho^{(\alpha)}_2(M)\|_1.
\label{CS-dist}
\end{equation}
In particular,
$D( C({\cal U}_1),\, C({\cal U}_2))=0$ iff $\forall M\subset \Lambda$
one has either $\varphi_{{\cal U}_1}(M)=\varphi_{{\cal U}_2}(M)\neq 0$ and 
$\rho^\alpha_{{\cal U}_1}(M)=\rho^\alpha_{{\cal U}_2}(M),\,(\alpha=c,e)$, 
or $\varphi_{{\cal U}_1}(M)=\varphi_{{\cal U}_2}(M)=0$.
In words: two conceptual structures are the same iff all the core
effects/causes repertoires and the associated integrated-information
coincide for all concepts.

It is important to notice that if the repertoires depend continuously
on some parameter, e.g., through the map $\cal U$, then $\varphi(M)$
will be a continuous function as well.
However, core effects/causes may change dis-continuously and this will
be reflected by CS operators (\ref{CS}) and functions thereof, e.g., Eq.~(\ref{CS-dist}).
\vskip 0.2truecm
{\bf{C: The CS of the Swap}}
The network supports just two concepts. The conceptual structure operator  is given bu: $C(S) =\frac{1}{2}\sum_{\alpha=e,c}\left( \frac{1}{2} |1\alpha\rangle\langle 1\alpha|\otimes (  \frac{ {\mathbf{1}} }{2} \otimes \Psi )+
\frac{1}{2} |2\alpha\rangle \langle2\alpha|\otimes ( \Psi\otimes  \frac{ {\mathbf{1}} }{2}) \right).$ 
The core effect/cause of  $M=\{1\}$ ($M=\{2\}$) is  $P=\{2\}$ ( $P=\{1\}$).
\vskip 0.2truecm
{{The key idea in IIT is to compare the global cause/effect structure (encoded in our quantum version in (\ref{CS})) with those of factorized maps
associated to  bi-partitioned and decoupled networks. In this way one wants to assess how the ``whole goes beyond and above the sum of its parts'' i.e., it exists intrinsically}}
The standard way in classical IIT to produce factorized maps is by
bi-partitioning the total set $\Lambda$ and by ``cutting the
connections between the two halves by injecting them with
noise"~\cite{IIT-2}.
We adopt here a natural quantum version of this procedure. 
Given the (non-trivial) partition
${\cal P}= \{\Lambda_1, \Lambda_2=\Lambda_1^\prime\}$, one can define 
\begin{equation}
{U}_{\cal P}= {\cal U}_1 \otimes {\cal U}_2,\quad
{\cal U}_i \colon L({\cal H}_{\Lambda_i}) \rightarrow L({\cal H}_{\Lambda_i})\colon X\mapsto {\cal U}_i (X) :={\mathrm{Tr}}_{\Lambda_i^\prime}  {\cal U}(X\otimes \frac{ {\mathbf{1}}_{\Lambda_i^\prime} } {d^{|\Lambda_i^\prime|}}),\quad (i=1,2)
\label{factorized-I}
\end{equation}
Notice that the ${\cal U}_i$'s, while unital, are not in general unitary even if the unpartitioned map $\cal U$ is.
We are now finally ready to define the fundamental global quantity of the paper: the
Integrated Information, denoted by $\Phi,$ of the whole network. Qualitatively, $\Phi$ measures how the integrated cause/effect
structure of the  quantum network fails to be described by any partitioned and decoupled version of it.
\vskip 0.2truecm
{\bf{Definition 5: Integrated Information.-- }}
We define Quantum {{Integrated Information}} (II) by 
\begin{equation}
\Phi({\cal U}):=\min_{{\cal P}} D(C({\cal U}), C({{\cal U}_{\cal P}})). \label{Phi}
\end{equation}
The minimum here is taken over the set of $2^{|\Lambda|-1}-1$ bi-partitions of $\Lambda.$
If $\Phi({\cal U})=0$, we say that the  network  $(\Lambda, {\cal U}, \Psi_\Lambda)$
is {\em{dis-integrated}}. The bi-partition ${\cal P}_{MIP}$, for which the minimum in
Eq.~(\ref{Phi}) occurs, is referred to as the Maximally
Irreducible Partition (MIP) in classical IIT, i.e.,
$\Phi({\cal U})=D(C({\cal U}), C({\cal U}_{  {\cal P}_{MIP}}))$.
\vskip 0.2truecm
 If the network is dis-integrated 
$C({\cal U})=C({\cal U}_{{\cal P}_{MIP}})$, namely there exists a
``cut and noising'' of the network in two halves that
does not affect its global (integrated) cause/effect structure. 
The system does not exist as a whole {\em{per se}}; 
in a network-intrinsic  information-theoretic sense there is no ``added value" in combining the two halves. 
For a completely factorized $\Psi_\Lambda$ and  ${\cal U}={\mathbf{1}}$ one has  ${\cal U}_{\cal P}={\cal U},\forall{\cal P}\Rightarrow \Phi({\mathbf{1}})=0.$
\vskip 0.2truecm
{\bf{D: $\Phi$ of the Swap}} We have of course just one partition which dis-integrates  both concepts in $C(S).$ Therefore, using (\ref{CS-dist}) and (\ref{Phi}) one has 
$$\Phi(S)=2\times \frac{1}{4}\left(  \| \frac{1}{2} ( \frac{  {\mathbf{1}} }{2}\otimes\Psi)-0\|_1 +  \| \frac{1}{2} ( \Psi\otimes \frac{  {\mathbf{1}} }{2})-0\|_1\right)=
 \frac{1}{2} ( \frac{1}{2}+\frac{1}{2})=\frac{1}{2}.$$
\vskip 0.2truecm
Several remarks 
 are now in order to shed some light on the nature of the quantum II  defined by Eq.~(\ref{Phi}).

{\bf{7)}}  $\Phi$ obeys ``time-reversal symmetry" $\Phi({\cal U}^*)=\Phi({\cal U})$~\cite{time-reversal}  and, for unitary ${\cal U}_i$'s,  $\Phi( (\otimes_{i\in\Lambda}{\cal U}_i){\cal U})=\Phi({\cal U})$~\cite{group-action}.

{\bf{8)}} In spite of the simplified notation, one should not forget that the conceptual structure operator and, therefore $\Phi$ depends on $\Psi_\Lambda$ as well.
In this paper we will focus at first on {\em{completely factorized pure states}} 
$\Psi_\Lambda = \otimes_{i\in\Lambda} |\psi_i\rangle\langle\psi_i|$.
In this case factorizability of ${\cal U}$ is a sufficient   condition for vanishing $\Phi.$ In fact, for a {\em{given}} $\Psi_\Lambda$, vanishing $\Phi$
is a {\em{weaker}} property than factorizability of the dynamical map.  Take, e.g., any non-factorizable unitary ${\cal U}$ that is diagonal
in a tensor product basis and $\Psi_\Lambda$ to be any basis element.
One has that ${\cal U}(\Psi_M\otimes \frac{ {\mathbf{1}}_{M^\prime}   } {d^{ |M^\prime|} })=  \Psi_M\otimes \frac{ {\mathbf{1}}_{M^\prime} }{d^{|M^\prime|}},\,(\forall M\subset \Lambda)$.
The action of $\cal U$, for this $\Psi_\Lambda$, is the same of the
identity map and therefore $\Phi({\cal U})=0$.

{\bf{9)}} It is essential  to stress  that different state choices for $\Psi_\Lambda$, e.g., entangled, may result in dramatically
different result. For example, {\em{even factorized maps  may have non-vanishing $\Phi$}}. 
To illustrate this intriguing fact let us consider e.g., $d=2,\,|\Lambda|=2,\,{\cal U}={\mathbf{1}},\,|\Psi_\Lambda\rangle=\frac{1}{\sqrt{2}}(|00\rangle+|11\rangle).$
One can easily see   that  there is just one concept (supported by the full $\Lambda$) and one partition ${\cal P}=\{\Lambda_1=\{1\},\,\Lambda_2=\{2\}\},$ from which it follows   that $\Phi({\mathbf{1}})=\frac{3}{4}>0$ \cite{unp}. This is an example of what might be dubbed {\em{ entanglement activated integration}}, and it  shows a sense in which genuinely quantum effects may play a ``positive" role in our version of IIT.

{\bf{10)}} At the quantum level one might define the minimization (\ref{Phi})
over all possible {\em{virtual}} bi-partitions of
${\cal H}_\Lambda$~\cite{virtual1,virtual2}. This would provide a lower bound to $\Phi$ and a much more stringent,
and uniquely quantum, definition of integration. Of course at the computational level this would be a tremendous challenge.

{\bf{11)}} If $\Omega\subset \Lambda$ one can consider the reduced network $( {\cal H}_\Omega, {\cal U}_\Omega, \Psi_\Omega)$ where:
if $X\in L({\cal H}_\Omega)$ then ${\cal U}_\Omega (X):={\mathrm{Tr}}_{\Omega^\prime} {\cal U}(X\otimes \frac{{\mathbf{1}}_{\Omega^\prime}}{d^{|{\Omega^\prime}|}}),$ and 
$\Psi_\Omega ={\mathrm{Tr}}_{\Omega^\prime}\Psi_\Lambda.$ 
If $\tilde{\Omega}\cap{\Omega}\neq\emptyset\Rightarrow \Phi({\cal U}_\Omega)\ge \Phi({\cal U}_{\tilde{\Omega}})$
the reduced network is referred to as a {\em{complex}} \cite{IIT-0,IIT-1}.  
A network may have many complexes which  represent, in a sense, ``local maxima" of
 $\Phi.$

\vskip 0.2truecm
We are now  ready to illustrate the  rather complex mathematical framework developed so far  by means of  physically motivated examples. 
Let us start with a very simple one.
\vskip 0.2truecm
\subsection{Partial Swap}
\begin{figure}[h!]
\includegraphics[scale=0.75]{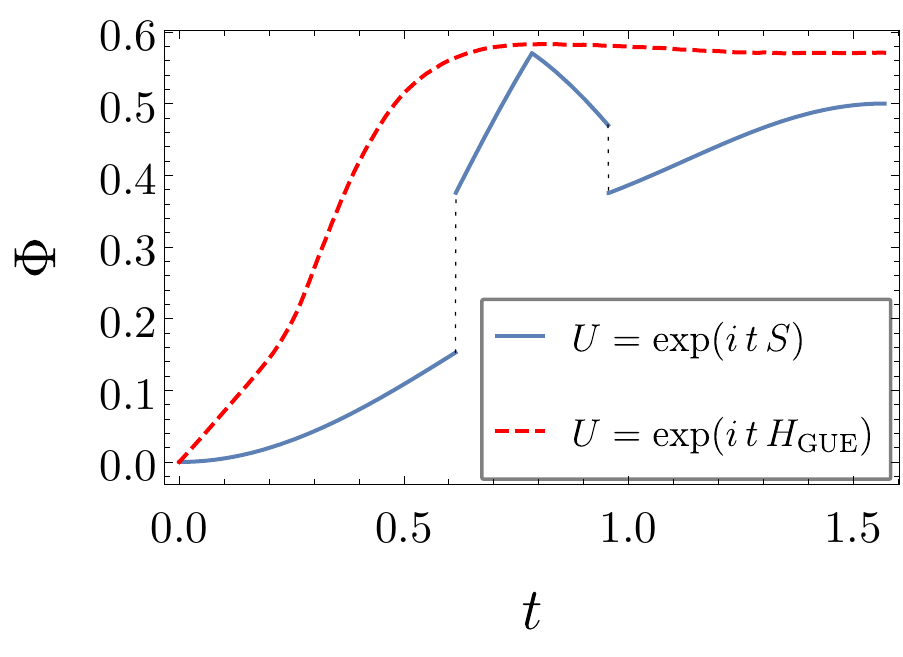}
\caption{ {\small{
   Two qubit network. Solid blue curve: $U=\exp(i\,t\,S)$. For small
  $t$ (near $\frac{\pi}{2}$) the network is in the ``identity''
  (``swap'') phase.
  The discontinuities at $t=\cos^{-1}(\sqrt{2/3})$ and
  $t=\cos^{-1}(1/\sqrt{3})$ are due to a jump and delocalization of the
  core cause/effect repertoires.
  The dashed red curve shows the average of $\Phi$ where
  $U=\exp(i\,t\,H_{\mathrm{GUE}})$ and $H_{\mathrm{GUE}}$ is sampled
  from the Gaussian unitary ensemble (GUE) with unit variance. }}  
}
\label{fig:PHItwoq}
\end{figure}

Let us consider a basic network with 
$|\Lambda|=2,\,d=2,\,\Psi_\Lambda=\Psi^{\otimes\,2}$
($\Psi$ pure state paperion), 
equipped with a ``partial swap" map
${\cal U}_t(X):= e^{it S}Xe^{-it S},\,(t\in[ 0,\frac{\pi}{2}]).$
%
One has three mechanisms/purviews $M, P=\{1\},\{2\},\Lambda=\{1,2\}$.
Direct computation 
shows~\cite{unp}: 
\begin{eqnarray}
\rho^{(e/c)}(1|1)&=&c_t^2 \Psi+s_t^2 {\mathbf{1}},\quad
 \rho^{(e/c)}(2|1)=s_t^2 \Psi +c_t^2 {\mathbf{1}}, 
\nonumber\\
\rho^{(e/c)}(\Lambda|1)&=&c_t^2 \Psi \otimes \frac{{\mathbf{1}}}{2} +s_t^2  \frac{{\mathbf{1}} } {2}\otimes \Psi  \pm
ic_t s_t [S, \Psi \otimes \frac{{\mathbf{1}}}{2}].
\label{S-reps}
\end{eqnarray}
Identical expressions hold for the repertoires $\rho^{(x)}(P|2),$
($c_t=\cos t,\,s_t=\sin t$).
Finally, $\rho(1|\Lambda)=\rho(2|\Lambda)=\Psi $ and $\rho^{(x)}(\Lambda|\Lambda)=\Psi ^{\otimes\,2}.$
One can obtain $\varphi$ for each mechanism $(i=1,2)$
$\varphi^{e/c}_t(i)=\frac{1}{2} \max \big\{ c_t^2, s_t^2,\min\{ \frac{s_t^2}{2}+\sqrt{ (\frac{s_t^2}{2})^2+   c_t^2 s_t^2   },    \frac{c_t^2}{2}+\sqrt{ (\frac{c_t^2}{2})^2+   c_t^2 s_t^2   } \} \big\},$ and $\varphi^{e/c}_t(\Lambda)=\frac{1}{2} \min\{ s_t^2(2-\frac{s_t^2}{2}),\,  c_t^2(2-\frac{c_t^2}{2})   \}$.
It follows that for small (near to $\frac{\pi}{2}$) $t$'s the core
effect/cause of $\{1\}$  is itself ($\{2\}$) (analogously for
$\{2\}$), whereas the core effect/cause of $\Lambda$ is itself
$\forall t$.
Moreover, there is a window around $t=\frac{\pi}{4}$ in which the core
effect/cause of $\{1\}$ and $\{2\}$ delocalize and comprise the full
$\Lambda$.
The corresponding  jumps 
of $\Phi(t)$ are shown in
Fig.~\ref{fig:PHItwoq}.
The  intermediate, high $\Phi$, delocalized phase
originates from the commutator term in (\ref{S-reps}).
It can be regarded as a genuine quantum feature, i.e., it would
disappear if ${\cal U}_t$  were just a probabilistic mixture of
identity and swap.
%
%
\subsection{Permutational networks}
\begin{figure}[h!]
\includegraphics[scale=0.25]{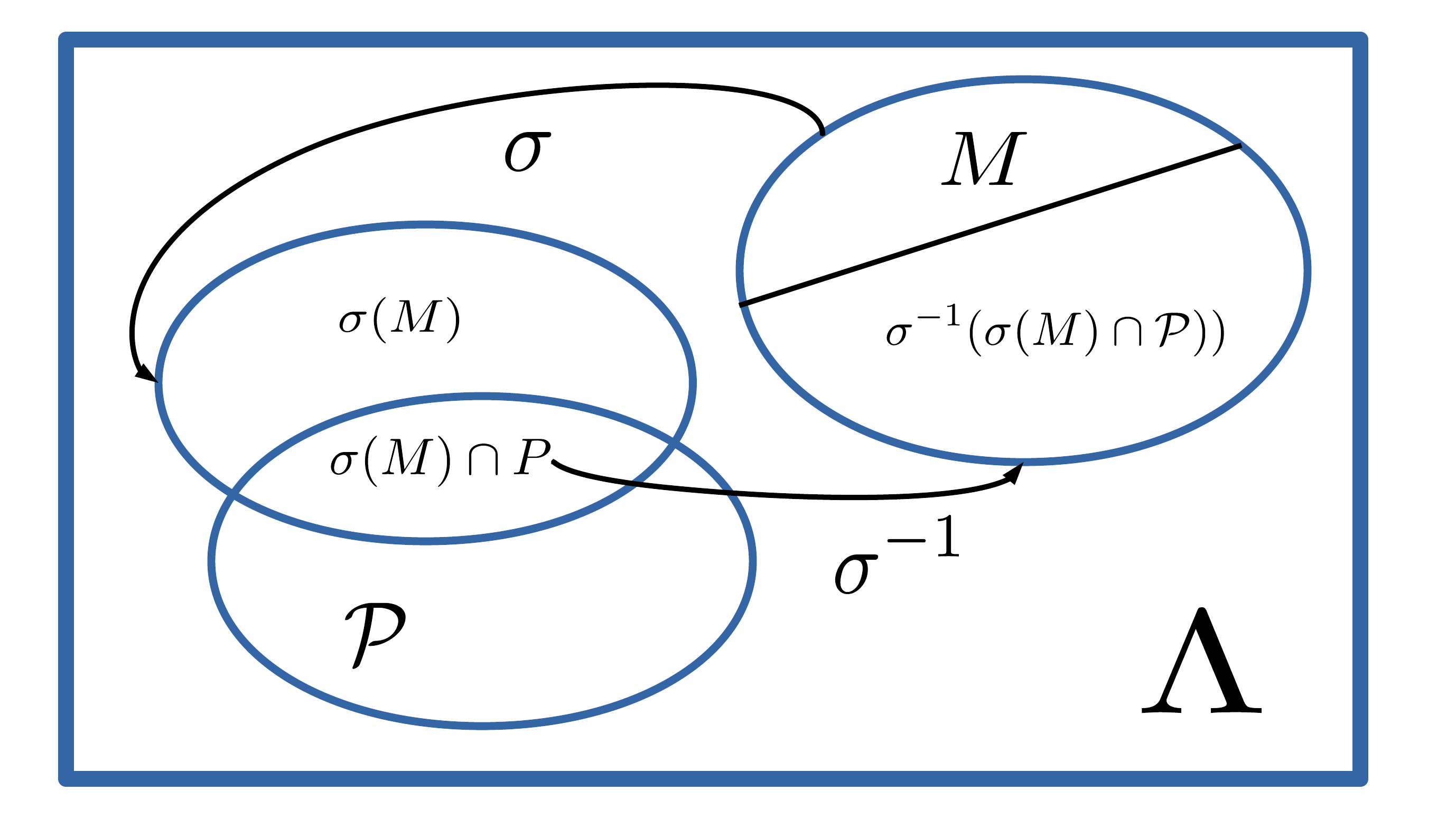}
\caption{{\small{ Permutational networks}}}
\label{fig:perm-net-fig}
\end{figure}
\begin{figure}[h!]
\includegraphics[scale=0.3]{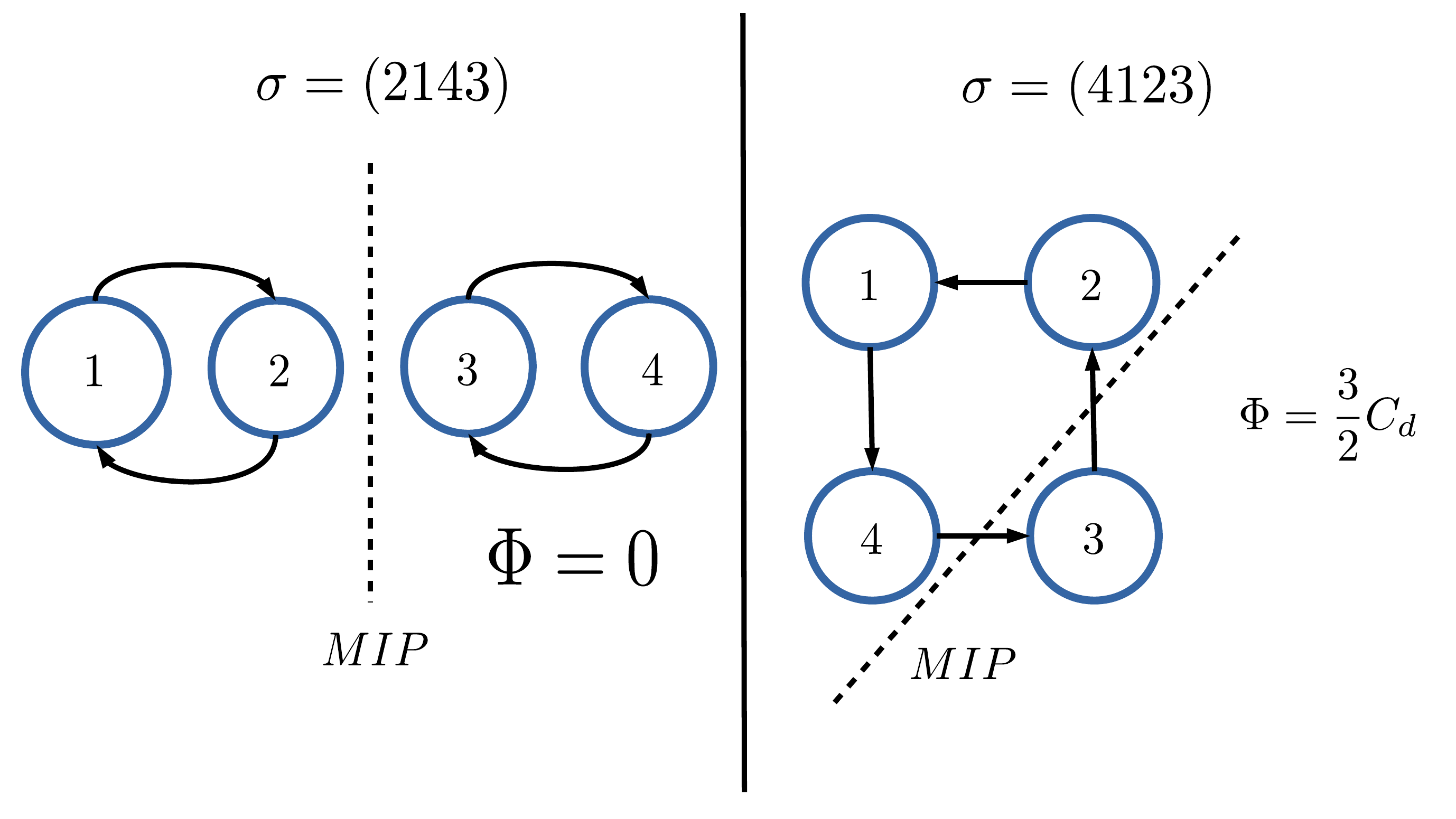}
\caption{ {\small{
On the left: a permutation in ${\cal S}_4$ which factorizes in two disjoint (order two) cycles. The MIP does not  dis-integrate any concept and this results in $\Phi=0.$ On the right:  an order four cycle. Here the MIP  dis-integrate three concepts: the core effect (cause) of site $4$ ($2$) is on the other side of the cut, while both core cause and effect of site $3$ are on the other side. This results (see text) in $\Phi=\frac{3}{2} c_d.$ }}
  }
\label{fig:perm-nets}
\end{figure}

Let us now discuss the  case of {{Permutational networks}} which is another obvious generalization of the  Swap case.  Here
$\Psi_\Lambda=\otimes_{i\in\Lambda}|\psi_i\rangle\langle\psi_i|$ 
and
${\cal U}(X)=U_\sigma X U_\sigma^\dagger$, where $U_\sigma$ acts as
the permutation  
$\sigma\in{\cal S}_{|\Lambda|}$ over
${\cal H}_\Lambda\cong ({\mathbf{C}}^d)^{\otimes\,|\Lambda|}$ i.e., $U_\sigma \otimes_{i\in\Lambda}|\psi_i\rangle=\otimes_{i\in\Lambda}|\psi_{\sigma(i)}\rangle.$
One can see that~\cite{unp}
\begin{eqnarray}
&&\rho^{(e)}(P|M)=\Psi_{\sigma^{-1}({P\cap \sigma(M))}}\otimes \frac{ {\mathbf{1}}_{P\cap \sigma(M)^\prime}} {d^{|P\cap \sigma(M)^\prime|}}=\rho^{(e)}(\sigma(M)\cap P|M)\otimes \rho^{(e)}(\sigma(M)^\prime \cap P|\emptyset)\nonumber\\
&&\rho^{(e)}(\sigma(M)\cap P|M)=\otimes_{j\in \sigma^{-1}(P)\cap M}
\rho^{(e)}(\sigma(j)|j) \otimes \rho^{(e)}(\emptyset|\sigma^{-1}(P^\prime)\cap M).
\end{eqnarray}
(See Fig.~(\ref{fig:perm-net-fig})). The same equations hold, with $\sigma^{-1}$ replacing $\sigma$, for
the cause repertoires.
From this totally factorized form one sees that the only irreducible
$(M,P)$ pairs are given by $(i, \sigma^{\pm1}(i))$
[and that the core effect (cause) of $i\in\Lambda$ is $\sigma(i)$
($\sigma^{-1}(i)$)] 
with $\varphi(i)= 1-d^{-1}=:c_d$.
From these results and (\ref{CS}) one has 
\begin{equation}
C(\sigma)=\frac{c_d}{2}\sum_{i\in\Lambda,x=\pm 1} |i x\rangle\langle i x |\otimes\rho^{(x)}(i),
\qquad 
\rho^{(x)}(i):=|\psi_{\sigma^x(i)}\rangle\langle \psi_{\sigma^x(i)}|\otimes \frac{ {\mathbf{1}}_{\{\sigma^x(i)\}^\prime}}{d^{|\Lambda|-1}}.
\label{CS-perm}
\end{equation}
Now given the partition
${\cal P}=(\Omega,\Omega^\prime),\,(\Omega\neq \emptyset)$,
from the {{Dis-integration Lemma}} in the Appendix   it follows that the concepts which
are dis-integrated are those whose core effects or/and causes lie on
the complementary set.
Any permutation can be factorized in disjoint cycles, if the number of
cycles is larger than one,
one can choose a partition the of $\Lambda$ that gives rise to the
same CS and, therefore $\Phi(\sigma)=0$. Each of the cycles will give rise to a complex with locally maximum $\Phi.$
If there is just one cycle (and $|\Lambda|>2$) the MIP is anyone of
the form ${\cal P} =\{ \{i\}, \{i\}^\prime\}$ in which the three
concepts associated with $i,\sigma(i),\sigma^{-1}(i)$ are
dis-integrated and all the others left intact (see Fig.~(\ref{fig:perm-nets})).
It follows from (\ref{Phi}) that
$\Phi(\sigma)= \frac{3}{2} c_d=O(1)$.
For $|\Lambda|=2$, just two concepts are dis-integrated by the only
possible partition and $\Phi(\sigma)=c_d$.
%
\section{Holistic and low-integration phases}
As customary in statistical mechanics one can consider families of
increasingly large networks $(\Lambda, {\cal U}_\Lambda,\Psi_\Lambda)$
and study how $\Phi$ behaves in the ``thermodynamical limit" (TDL)
$|\Lambda|\to\infty.$ If the maps ${\cal U}_\Lambda$ are associated with unitaries
$U_\Lambda=e^{-it_\Lambda H_\Lambda}$ one has to choose how to scale
with  $|\Lambda|$ both the times $t_\Lambda$ as well as
the Hamiltonians $H_\Lambda$.
For the former three natural options are: {\em{a)}} $t=O(1)$;
{\em{b)}} ``constant action" $t_\Lambda \|H_\Lambda\|=O(1)$; {\em{c)}}
$t_\Lambda:={\mathrm{argmax}}_t \Phi_\Lambda(t)$, where the maximum
over $t$ is taken at fixed $|\Lambda|$ 
\vskip 0.2truecm
{\bf{Definition 6: Holistic Phases}} When
\begin{equation}
\mathcal{O}({\cal U}):=\lim_{|\Lambda|\to\infty} \frac{\log_2\Phi({\cal U}_\Lambda)}{|\Lambda|}>0,
\label{holo-param}
\end{equation}
we say that the network $(\Lambda, {\cal U}, \Psi_\Lambda)$ is the {\em{holistic phase}} in the TDL.
\vskip 0.2truecm
In the holistic phase the system shows the maximal level of integration as an irreducible causal whole.
The quantity $\mathcal{O}({\cal U})$ can be referred to as the holistic
parameter and it is at most one \cite{at-most-one}

%
To study the different integration phases 
it is useful to consider the following upper bound to  $\Phi$~\cite{upper-bound}
\begin{equation}
\Phi({\cal U})\le {\mathrm{Tr}}\, C({\cal U})=\sum_{M\subset\Lambda}\varphi_{\cal U}(M) \le N_c({\cal U}),
\label{upper-bound-Nc}
\end{equation}
\vskip 0.2truecm

If $ N_c({\cal U})=O( |\Lambda|^\kappa)$ with $\kappa >0$ then,
from (\ref{upper-bound-Nc}), it follows that the holistic parameter 
is vanishing:
$\mathcal{O}({\cal U})=\lim_{|\Lambda|\to\infty}|\Lambda|^{-1}\log_{2}\Phi({\cal U})\le \lim_{|\Lambda|\to\infty}|\Lambda |^{-1}\log_{2} |\Lambda|^\kappa=0$. 
On the other hand, in order to be in the holistic phase, the network
needs to have a number of concepts asymptotically lower bounded by
$2^{a |\Lambda|}$ ($1 \geq a > 0$).
Whence, if the concepts are supported only on mechanisms $M$, such
that $|M|=\kappa=O(1)$, then the network is necessarily in the
non-holistic  phase. The permutational case discussed in the former section, where concepts are supported by sites of $\Lambda$ only,  provides an example of these low $\Phi$ networks.
\subsection{Holistic Phase}
%
In this section we  discuss a sufficient condition for a network to be in the
holistic phase and provide a physical example. 
Before doing so we need to define  the ``boundary of the partition"
Given a (non-trivial) bi-partition ${\cal P}:=\{\Lambda_1,\,\Lambda_2=\Lambda_1^\prime\}$
we define $\partial {\cal P}:=\{ S\subset \Lambda\,/\, S\cap\Lambda_1\neq \emptyset \wedge  S\cap\Lambda_2\neq \emptyset \}.$
This set contains $|\partial{\cal P}|=2^{|\Lambda|}-2^{|\Lambda_1|}-2^{|\Lambda_2|}+1$ elements.   
Now, one can prove that~\cite{lower-bound}
\begin{equation}
\Phi({\cal U})\ge \frac{1}{2}\sum_{M\in\partial{{\cal P}_{MIP}} }\varphi_{\cal U}(M)\ge |\partial{\cal P}_{MIP}|\frac{\varphi_0}{2},\label{Phi-lower-bound}
\end{equation}
where $\varphi_0:=\min_{M \in\partial{{\cal P}_{MIP}} } \varphi_{\cal U}(M).$ 
Notice that $|\partial{\cal P}_{MIP}|\ge 2^{|\Lambda|-1}-1$, therefore, in view of the lower bound above, one is guaranteed to be
in the holistic phase if $\varphi_0$ is lower-bounded by a non-zero constant.
\begin{figure}[h!]
 \begin{center}
  \begin{tabular}{cc}
   \includegraphics[scale=0.7]{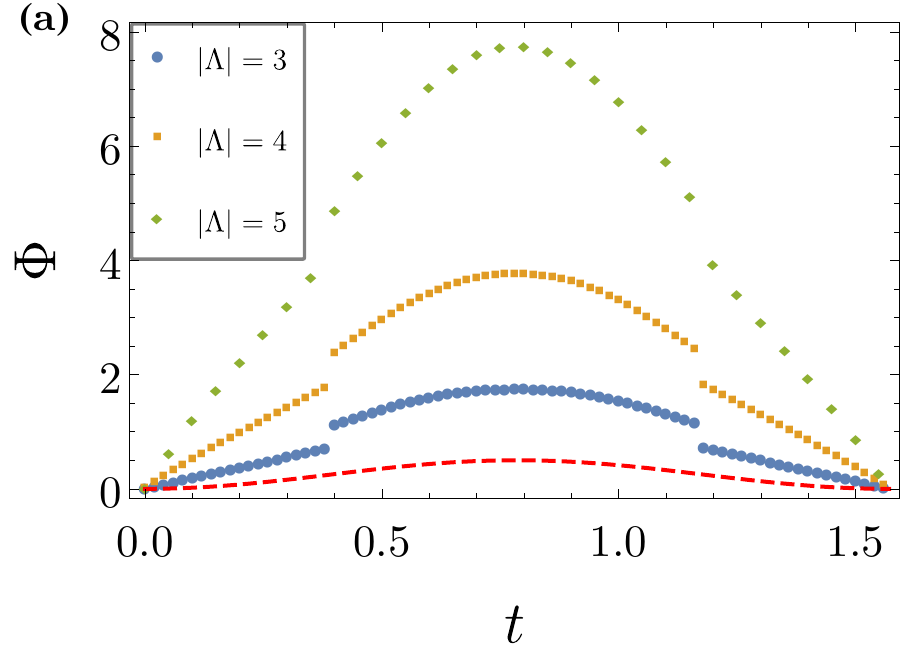}
   &
   \includegraphics[scale=0.7]{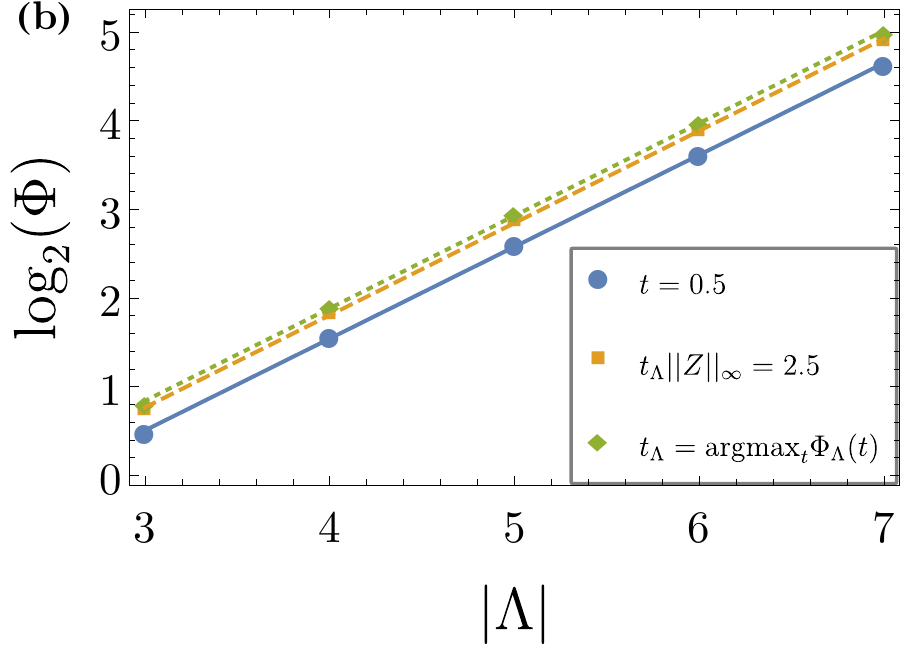}
  \end{tabular}
  \caption{ {\small{
  (a)   $\Phi$ as a
  function of the parameter $t$, for different system sizes $|\Lambda|=3,4,5$,
  with $\Psi_\Lambda = \otimes_{i\in\Lambda}|+\rangle\langle +|_i$.
  The red dashed line shows $2s_t^2 c_t^2$.
  (b) Scaling of $\Phi$ with the system size $|\Lambda|$, for the
  three options to fix the time scale {\it a)} $t=0.5$ (solid blue line)
  $\log_{2}\Phi=1.04 |\Lambda|-2.60$, {\it b)}
  $t\|Z\|_{\infty}=2.5$ (dashed orange line)
  $\log_{2}\Phi=1.04 |\Lambda|-2.35$ and {\it c)}
  $t=\mathrm{argmax}_{t}\Phi_{\Lambda}(t)$ (dotted green
  line) $\log_{2}\Phi=1.04 |\Lambda|-2.29$.}}
  }
 \label{fig:ZPHI}
\end{center}
\end{figure}

{\bf{Example: $|\Lambda|$-local interaction: }}
Let us consider a qubit network of size $|\Lambda|$ with 
\begin{equation}
{\cal U}(X)=e^{itZ}Xe^{-itZ},\,Z:=\otimes_{i\in\Lambda}\sigma_i^z,\quad\Psi_\Lambda= \otimes_{i\in\Lambda}|+\rangle\langle +|_i=:\Pi^{+}_\Lambda.
\end{equation}
In the Appendix is shown that, for $M\neq \Lambda,$ one has
$\varphi(M|M)=2s_t^2 c_t^2.$ 
Whereas, for the case $M=\Lambda$  one finds  $\varphi(\Lambda|\Lambda)=| s_t c_t| (1+|s_t c_t|)\ge 2s_t^2 c_t^2$.
Since, by definition, $\varphi_{{\cal U}_t}(M)\ge \varphi_{{\cal U}_t}(M|M)$ by setting, e.g., $t=\frac{\pi}{4}$, one finds 
$\varphi_0\ge \frac{1}{2}$. This also shows that the holistic parameter $\mathcal{O}({\cal U}_t)$
is one for all $t\neq 0, \pi/2$, where it is ill-defined as $\Phi=0$. 
Turning on the global interaction $Z$ (or mixing it with the $\mathbf{1}$) results in a direct transition from the dis-integrated phase to the holistic one.  

In Fig.~\ref{fig:ZPHI}~(a) we plot $\Phi(t)$, for different system
sizes $|\Lambda|=3,4,5$, given the initial state
$\Psi_\Lambda = \otimes_{i\in\Lambda}|+\rangle\langle +|_i$.
As predicted by the above analytical calculations, we find
$\Phi>2s_t^2 c_t^2$, (the dashed red line represents $2s_t^2 c_t^2$), 
and for $t=0$ and $t=\pi/2$ we obtain $\Phi=0$.
Fig.~\ref{fig:ZPHI}~(b) shows the holistic behaviour of the network,
i.e., the exponential scaling of $\Phi$ with the system size
$|\Lambda|$.
In particular, we observe an exponential scaling of $\Phi$ for all the three
natural prescriptions for fixing the timescale:
{\it a)} for $t=0.5$ (solid blue line) we get $\log_{2}\Phi = 1.05 |\Lambda| - 2.64$,
{\it b)} for $t \|Z\|_{\infty}=2.5$ (dashed orange line),
$\log_{2}\Phi = 1.04 |\Lambda| - 2.35$ and
{\it c)} for $t=\mathrm{argmax}_{t}\Phi_{\Lambda}(t)$
(dotted green line), $\log_{2}\Phi=1.04 |\Lambda|-2.29$. The fits
obtained using all the three different prescriptions are consistent
with a scaling of the form $\Phi \sim 2^{|\Lambda|}$.
%
%

%
\subsection{Low-integration: $O(1)$-local interactions}
\begin{figure}[h!]
\includegraphics[scale=0.75]{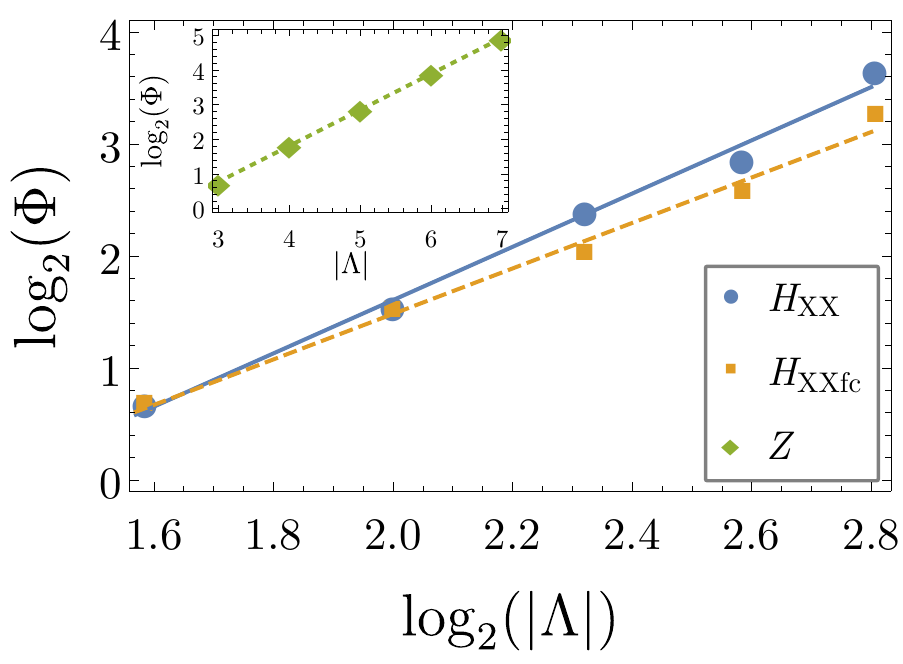}
\caption{
{\small{ 
  Low-integrated and holistic networks:
  Numerical simulations of $\Phi$, for the two-body Hamiltonian on a
  ring
  $H_{\mathrm{XX}} = \sum_{i=1}^{|{\Lambda}|}(\sigma_{i}^{x}\sigma_{i+1}^{x}+\sigma_{i}^{y}\sigma_{i+1}^{y})$, 
  and on a fully connected graph $H_{\mathrm{XXfc}}=\sum_{i<j}^{|{\Lambda}|}(\sigma_{i}^{x}\sigma_{j}^{x}+\sigma_{i}^{y}\sigma_{j}^{y})$,
  as well as for a $|\Lambda|$-local interaction
  $Z=\bigotimes_{i=1}^{|{\Lambda}|}\sigma_{i}^{z}$, are depicted. 
  %
  For $H_{\mathrm{XX}}$ ($H_{\mathrm{XXfc}}$) the fit gives
  $\log_{2}(\Phi)=2.38\log_{2}|{\Lambda}|-3.15$
  ($\log_{2}(\Phi)=2.03\log_{2}|{\Lambda}|-2.58$), 
  illustrating a polynomial growth of the integrated information $\Phi$
  with the system size $|{\Lambda}|$ (low-integration).
  The same behavior is also observed for the Heisenberg XXX model.
  Inset: holistic phase for the $|\Lambda|$-local interaction $Z$.
  The fit shows $\log_{2}(\Phi)=1.04 |{\Lambda}|-2.35$, consistent
  with an exponential scaling $\Phi\sim 2^{|\Lambda|}$.
%
  %
  The time-step $t$ for the different system sizes $|{\Lambda}|$ is
  fixed by the ``constant action" prescription $t
  \|{H}\|_{\infty}=2.5$.
  %
}}  
  }
\label{fig:PHInetworks}
\end{figure}

%
In general, using Lieb-Robinson type arguments, one might be tempted to  speculate that $k$-local ($k=O(1)$) interactions
will give rise to low-integration networks with sub-extensive
$\log_{2}\Phi$~\cite{unp}. 
Preliminary numerical results are shown
in Fig.~\ref{fig:PHInetworks} in which 
${\cal U}(X)=e^{-it_{\Lambda}H}Xe^{it_{\Lambda}H}$,
but now the dynamics is generated by
a two-body Hamiltonian $H$.
Namely, the dynamics are governed by i) the XX Hamiltonian on a ring
$$
H_{\mathrm{XX}}
=
\sum_{i=1}^{|{\Lambda}|}
(\sigma_{i}^{x}\sigma_{i+1}^{x}+\sigma_{i}^{y}\sigma_{i+1}^{y}),
$$
ii) by the XX Hamiltonian on a fully connected graph 
$$
H_{\mathrm{XXfc}}
=
\sum_{i<j}^{|{\Lambda}|}
(\sigma_{i}^{x}\sigma_{j}^{x}+\sigma_{i}^{y}\sigma_{j}^{y}),
$$
iii) by the XXX Hamiltonian on a ring
$$
H_{\mathrm{XXX}}
=
\sum_{i=1}^{|{\Lambda}|}
(\sigma_{i}^{x}\sigma_{i+1}^{x}+\sigma_{i}^{y}\sigma_{i+1}^{y}+\sigma_{i}^{z}\sigma_{i+1}^{z}),
$$
and iv) by the XXX Hamiltonian on a fully connected graph
$$
H_{\mathrm{XXXfc}}
=
\sum_{i<j}^{|{\Lambda}|}
(\sigma_{i}^{x}\sigma_{j}^{x}+\sigma_{i}^{y}\sigma_{j}^{y}+\sigma_{i}^{z}\sigma_{j}^{z}).
$$
%
The initial state was chosen to be
$
\Psi_\Lambda = \bigotimes_{i\in\Lambda} \ensuremath{\vert{0}\rangle}\ensuremath{\langle{0}\vert}_i
$,
whereas  for the holistic example with
$Z=\otimes_{i\in\Lambda}\sigma_i^z$,
we used the state
$\Psi_\Lambda=\bigotimes_{i\in\Lambda}|+\rangle\langle +|_i$, 
where
$|+\rangle=\frac{1}{\sqrt{2}}(|0\rangle+|1\rangle.$
\section{Conclusions}

The main goal of classical Integrated Information Theory (IIT)~\cite{IIT-0, IIT-1,IIT-2,IIT-3} is to provide a mathematical and conceptual framework to study the neural correlates of consciousness.
In this paper we took the first steps towards a possible quantum version of IIT {\em{irrespective}} of its applications to consciousness. 

Our approach is a quantum information-theoretic one, in which neural networks are being replaced by networks of qudits, probability distributions by non-commutative density matrices, and markov processes by completely positive maps. The irreducible cause/effect structure of the global network is encoded by a so-called conceptual structure operator. The minimal distance of the latter from those obtained by factorized versions of the network, defines the quantum Integrated Information $\Phi$. 

We have studied quantum effects in small qubit networks and provided examples, analytical and numerical, of families of low integration networks. Also, we have demonstrated sufficient conditions for the existence of highly integrated ones and given illustrations.

The scaling of $\Phi$ with the network size defines different phases distinguished by a different level of integration of their global cause/effect structure. 
The study of those phases and cross-overs, their relation to locality and entanglement, and in general the question whether the quantum IIT discussed in this paper has {\em{any}}
 direct bearing on standard quantum information processing, are challenging tasks for future investigations.

\emph{acknowledgements} P.Z. thanks the ``Waking up Podcast'' of Sam Harris for the inception, Giulio Tononi for useful input and  G. Styliaris for help with some of the pictures.
 Partial support from the NSF award PHY-1819189 is acknowledged.

\appendix
\section{Lieb-Robinson Bounds for Cause/Effect Information}
\label{sec:sup2}

We now assume that the CP map ${\cal U}$ is a unitary generated by a
{\em{local}} Hamiltonian $H_\Lambda$, i.e.,
${\cal U}(X)=e^{it H_\Lambda}Xe^{-it H_\Lambda}$. 
In this case one can show that a Lieb-Robinson type bound holds for
the cei (\ref{cei}).
Indeed,
\begin{eqnarray}
2\,ei(P|M)&=&\|\rho^{(e)}(P|M) -\frac{ {\mathbf{1}}_{P} } {d^{|P|}}\|_1={\mathrm{Tr}}_P \left(O_P ( \rho^{(x)}(P|M) -\frac{ {\mathbf{1}}_{P} } {d^{|P|}})\right) \nonumber \\
&=& {\mathrm{Tr}} \left( (O_P\otimes {\mathbf{1}}_{P^\prime}) {\cal U}( \Psi_M \otimes \frac{ {\mathbf{1}}_{M^\prime} } {d^{|M^\prime|}} -\frac{ {\mathbf{1}}_{\Lambda} } {d^{|\Lambda|}})\right) \nonumber \\
&=& {\mathrm{Tr}} \left( \tilde{O}_P (\Psi_M \otimes \frac{ {\mathbf{1}}_{M^\prime} } {d^{|M^\prime|}}-\frac{ {\mathbf{1}}_{\Lambda} } {d^{|\Lambda|}})\right),
\end{eqnarray} 
where $\tilde{O}_P:={\cal U}^*(O_P\otimes {\mathbf{1}}_{P^\prime})$.
Now one can write
$\Psi_M \otimes \frac{ {\mathbf{1}}_{M^\prime} } {d^{|M^\prime|}}=T_M(\frac{ {\mathbf{1}}_{\Lambda} } {d^{|\Lambda|}})$,
where $T_M$ is a ``preparation" CP Map, local to the $M$ mechanism
whose Kraus operators can be given by 
$A_k= |\Psi\rangle \langle k|\otimes \frac{ {\mathbf{1}}_{M^\prime} } {d^{|M^\prime|}}$ 
($\{|k\rangle\}_{k=1}^{d_M}$ is a basis for ${\cal H}_M$ and $\|A_k\|=1$).
Therefore,
\begin{align}
2\,ei(P|M)&={\mathrm{Tr}} \left( (T_M^*-{\mathbf{1}})(\tilde{O}_P)\frac{ {\mathbf{1}}_{\Lambda} } {d^{|\Lambda|}}\right)
\le \|(T_M^*-{\mathbf{1}})(\tilde{O}_P)\| \nonumber \\
&=\|T_M^*(\tilde{O}_P)-\tilde{O}_P\|= \|\sum_k (A_k^\dagger \tilde{O}_P A_k -A_k^\dagger A_k \tilde{O}_P\| \nonumber \\
&\le \sum_k \|A_k^\dagger\|\| [\tilde{O}_P, A_k]\|   
\le d_M \max_k  \| [\tilde{O}_P, A_k]\|.
\end{align}
Now, since $\tilde{O}_P$ is the evolution of an operator local at $P$
and the $A_k$'s are local to $M$, the Lieb-Robinson holds
in the form
\begin{equation}
ei(P|M)\le c \exp\left( -a (\mathrm{dist}(P,M)- v|t|)\right)=:\epsilon.
\label{LR-cei}
\end{equation}
Here $a, c>0$ are constants depending on $d_M, |M|, |P|,$
and $\|O_P\|$.
Moreover, $v>0$ is the Lieb-Robinson velocity, which depends on $H_\Lambda$.
Since ${\cal U}^*$ is also generated by a local Hamiltonian ($-H_\Lambda$),
an identical proof holds for $ci(P|M)$.
Finally, in the light of Eqs.~(\ref{upper-bound-ii}) and
(\ref{LR-cei}) one has that integrated-information fulfills a
Lieb-Robinson bound as well, i.e.,
$\varphi^{(e)}(P|M)\le c \exp\left( -a (\mathrm{dist}(P,M)- v|t|)\right)$ 
(and a similar one for $\varphi^{(c)}$).

From the Lieb-Robinson bound, by a standard argument, it also follows
that 
$\|\rho^{(x)}(\Lambda|M)- \rho^{(x)}(\tilde{M}^\prime|\emptyset)\otimes  \rho^{(x)}(\tilde{M}|M)\|\le \epsilon$,
where $\tilde{M} \supset M$ is a suitable Lieb-Robinson ``fattening"
of $M$.
From this inequality by taking traces with respect $P^\prime$, it
follows (again) that the repertoires $\rho^{(x)}(P|M)$ with purviews
$P\subset \tilde{M}$, are exponentially close to the unconditioned one
$\frac{\mathbf{1}}{d^{|P|}}$.
\section{Holistic phase example}
\label{sec:sup3}

Let us consider a qubit network of size $|\Lambda|$ with 
\begin{equation}
{\cal U}(X)=e^{itZ}Xe^{-itZ},\qquad Z:=\otimes_{i\in\Lambda}\sigma_i^z,\qquad \Psi_\Lambda= \otimes_{i\in\Lambda}|+\rangle\langle +|_i=:\Pi^{+}_\Lambda.
\end{equation}
For $M \neq \Lambda$, one directly finds
$\rho^{(e/c)}(P|M)= (c_t^2  \Pi^{+}_{M\cap P} +s_t^2 \Pi^{-}_{M\cap P})\otimes \frac{ {\mathbf{1}}_{P\cap M^\prime} } {2^{|P\cap M^\prime|} }$,
where $\Pi_X^{\pm}:=\otimes_{i\in X} |\pm\rangle\langle \pm|_i$
and $c_t:=\cos t,\,s_t:=\sin t$.
We now focus on the $P=M$ case with $|M|>1$ and look for
factorizations of the form
$\rho^{(e/c)}(\tilde{M}_1|M_1)\otimes \rho^{(e/c)}(\tilde{M}_2|M_2)$,
where $(M_1, M_2)$ and $(\tilde{M}_1, \tilde{M}_2)$ are (non-trivial)
pairings of $M$.
Using the above expression for the repertoires, one has that
$\rho^{(e/c)}(M|M)-\rho^{(e/c)}(\tilde{M}_1|M_1)\otimes \rho^{(e/c)}(\tilde{M}_2|M_2)$
is equal to
\begin{multline}
\Pi_{A}^+\otimes ( c_t^2   \Pi_{BC}^+- c_t^4  \frac{ {\mathbf{1}}_{BC} } {d_{BC} }   )\otimes \Pi_{D}^+
+\Pi_{A}^-\otimes ( s_t^2   \Pi_{BC}^-- s_t^4  \frac{ {\mathbf{1}}_{BC} } {d_{BC} }   )\otimes \Pi_{D}^- \\
-s_t^2c_t^2 (     \Pi_{A}^+\otimes\frac{ {\mathbf{1}}_{BC} } {d_{BC} }\otimes   \Pi_{D}^- + 
  \Pi_{A}^-\otimes\frac{ {\mathbf{1}}_{BC} } {d_{BC} }\otimes   \Pi_{D}^+),\nonumber
\end{multline}
where
$A:=M_1\cap \tilde{M_1}, B= \tilde{M}_1\cap {M_2}, C:= \tilde{M}_2\cap {M_1}, D:= M_2\cap \tilde{M_2}$,
and $d_{BC}:=2^{|B|+|C|}$.
The operator  above can be easily diagonalized giving
$$\sigma(X)=\{ 0, c_t^2 -\frac{c_t^4}{d_{BC}},  s_t^2 -\frac{s_t^4}{d_{BC}}, -\frac{c_t^2 s_t^2}{d_{BC}}, -\frac{c_t^4}{d_{BC}}, -\frac{s_t^4}{d_{BC}}\},$$
with degeneracies
$\{d^{|M|}-4d_{BC}, 1,1, 2d_{BC}, d_{BC}-1,d_{BC}-1\}$.
From this it follows
$\|X\|_1=2(1-\frac{c_t^4}{d_{BC}} -\frac{s_t^4}{d_{BC}})\ge 2(1-c_t^4-s_t^4)=4 s_t^2 c_t^2$.
The lower bound is achieved when $d_{BC}=1$, i.e., $M_1=\tilde{M}_1$
and $M_2=\tilde{M}_2$.
Finally,
$$\varphi(M|M)=\min_{(M_i,\tilde{M_i})} \frac{1}{2}\|\rho^{(e/c)}(P|M)-\rho^{(e/c)}(\tilde{M}_1|M_1)\otimes \rho^{(e/c)}(\tilde{M}_2|M_2)\|_1=
2s_t^2 c_t^2.$$
The case $M=\Lambda$ gives, with a similar calculation, $\varphi(\Lambda|\Lambda)=| s_t c_t| (1+|s_t c_t|)\ge 2s_t^2 c_t^2$.

\section{Computing $\Phi$}
Evaluating  quantum II, Eq.~(\ref{Phi}), involves several combinatorial layers and constitues (as in the classical case \cite{IIT-0}) a formidable computational challenge.
This implies that this paper entails  a non-negligible algorithmic component. We summarize here below our  strategy to compute  $\Phi:$
%
\vskip 0.2truecm
 {\bf{i)}} Compute $\rho^{(e/c)}(P|M)$ for all non empty $M,P\subset \Lambda$ (\# of repertoires:  $2\,(2^{2 |\Lambda|} -2^{|\Lambda|+1}+1)$)
%
 
{\bf{ii)}} For each non-trivial $\rho^{(e/c)}(P|M)$ compute $\varphi$ (\# of pairings of $(P,M)\colon$  $(2^{|M|+ |P|-1} -1)$)
%

  {\bf{iii)}} For each $M\neq \emptyset$ find its core effect/cause $P_*^{(x)}$ and associated integrated-information $\varphi^{(x)}(M)=\max_P \varphi^{(x)}(P|M)=\varphi^{(x)}(P_*^{(x)}|M),\,(x=e,c).$
Now the CS (\ref{CS}) is defined for the given ${\cal U}.$
%

 {\bf{iv)}} Iterate {\bf{i)}}--{\bf{iii)}} for each partition    $\cal P$ of $\Lambda$  (\# partitions of $\Lambda\colon$   $(2^{|\Lambda|-1} -1)$ to compute   $C({\cal U}_{\cal P})$ 

 {\bf{v)}} Compute $\Phi$ (Eq.~(\ref{Phi}) by finding the partition (MIP) which minimizes  $D(C({\cal U}),  C({\cal U}_{\cal P})$ (Eq.~\ref{CS-dist}).
\vskip 0.2truecm
The total number of steps (for a fixed partition) is $
\sum_{M,P\subset \Lambda } (2^{|M|+ |P|-1} -1)=\frac{1}{2}\sum_{|M|,|P|=0}^{|\Lambda|} {|\Lambda|\choose |M|}{|\Lambda|\choose |P|} (2^{|M|} 2^{|P|} -2)=
O({3^{2|\Lambda|}}/{2}).
$
As in the classical IIT case, the actual computation of $\Phi$ is
exponentially costly (in $|\Lambda|$) and therefore provides a
challenging task even for networks of moderate size.  
 
Here below state an important technical lemma (proven in the Appendix), which
shows how the CS is affected by partitioning the system (with  ${\cal P}=\{\Lambda_1,\,\Lambda_2=\Lambda_1^\prime\}$) and simplifies the algorithms for computing  $\Phi.$
\begin{figure}[h!]
\includegraphics[scale=0.25]{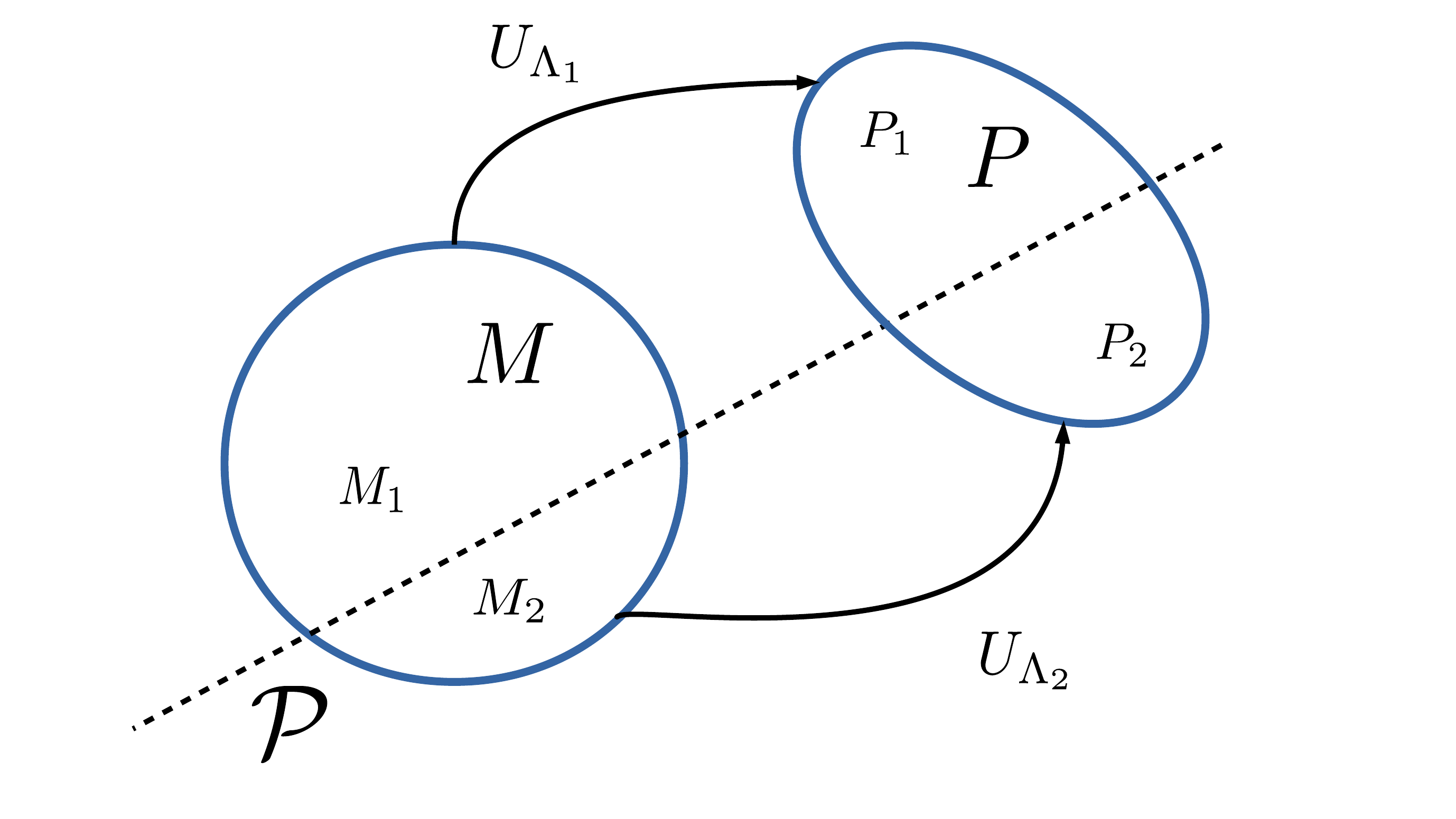}
\caption{ {\small{
 Illustration of the dis-integration lemma. The mechanism (purview) $M$ ($P$) is split in two sub-mechanism $M_1$ and $M_2$ ( $P_1$ and $P_2$) by the partition $\cal P.$ Since the ${\cal U}_{{\Lambda}_i}$'s ($i=1,2)$) connects only mechanisms on the ``same side" of $\cal P$ the conditional repertoires factorizes (see Eq.~(\ref{repertoires-factor})). }}
  }
\label{fig:dis-lemma}
\end{figure}
\vskip 0.2truecm
{\bf{Dis-integration Lemma}}
The cause/effect repertoires  of the partitioned map ${\cal U}_{\cal P}$ are factorized: 
\begin{equation}
\rho_{{\cal U}_{\cal P}}^{(e/c)}(P|M)=\rho_{\cal U}^{(e/c)}(P_1|M_1)\otimes \rho_{\cal U}^{(e/c)}(P_2|M_2),
\label{repertoires-factor}
\end{equation}
where  $P_i:=P\cap \Lambda_i,\, M_i:=M\cap\Lambda_i,\,(i=1,2)$ (see Fig.~(\ref{fig:dis-lemma}).
 \vskip 0.2truecm
From the factorized (\ref{repertoires-factor}) form it follows  that :  {\em{a) }} if both $M$ and $P$ are on the same side of the partition the $\rho^{(x)}(P|M)$ is unaffected 
 {\em{b)}}  If they are on opposite sides there is zero cause/effect information   ($ \rho^{(x)}(P|M)=\rho^{(x)}(P|\emptyset)\otimes \rho^{(x)}(\emptyset |M)=
 \frac{{\mathbf{1}}_P}{d^{|P|}}$)
{\em{c)}} if either one the two is straddling between the partition there is zero $\varphi(P|M).$
From a)--c) above one sees that {{ all the concepts such that $M\cup P_*^{(x)}\in \partial {\cal P}$
%
are dis-integrated whereas all those that $M\cup P_*^{(x)}\notin \partial {\cal P}$ are left invariant}}. 
From Eqs.~(\ref{CS-dist}) and (\ref{Phi}) we see that, for a given $\cal P$ just the former contributes to $\Phi$ (as the latter cancel being identical for $\cal U$ and ${\cal U}_{\cal P}$).
{{The MIP is then the partition that dis-integrates the least number of concepts in the CS of the undivided system.}} 
\vskip 0.2truecm
{\em{Proof.--}}
We define  $ \Psi_\Omega = \otimes_{i\in\Omega} \Psi_i,\,(\forall\Omega\subset\Lambda) .$
From Eqs  (\ref{repertoire}) and the definition of the factorized map one finds $\rho_{\cal P}^{(e/c)}(P|M)={\mathrm{Tr}}_{P^\prime}\, {\cal U}_{\cal P}\circ {\cal N}_{M^\prime}(\Psi_\Lambda),$
where  
\begin{equation}
{\cal U}_{\cal P}\circ {\cal N}_{M^\prime}(\Psi_\Lambda)=\otimes_{i=1}^2 {\cal U}_i \left(\Psi_{M_i}\otimes 
\frac{ {\mathbf{1}}_{M^\prime \cap\Lambda_i} } {d^{|M^\prime \cap\Lambda_i|}}\right) 
= \otimes_{i=1}^2 {\mathrm{Tr}}_{\Lambda_i^\prime}\, {\cal U} \left(\Psi_{M_i}\otimes 
\frac{ {\mathbf{1}}_{M^\prime \cap\Lambda_i} } {d^{|M^\prime \cap\Lambda_i|}}\otimes
\frac{ {\mathbf{1}}_{\Lambda_i\prime} } {d^{|\Lambda_i^\prime|}}  \right).\nonumber
\end{equation}
First notice that $\frac{ {\mathbf{1}}_{M^\prime \cap\Lambda_i} } {d^{|M^\prime \cap\Lambda_i|}}\otimes
\frac{ {\mathbf{1}}_{\Lambda_i\prime} } {d^{|\Lambda_i^\prime|}}= \frac{ {\mathbf{1}}_{M_i\prime} } {d^{|M_i^\prime|}}$
The result   follows  now from ${\mathrm{Tr}}_{P^\prime}= {\mathrm{Tr}}_{P^\prime\cap \Lambda_1}\otimes {\mathrm{Tr}}_{P^\prime\cap \Lambda_2}$ and ${\mathrm{Tr}}_{P^\prime\cap \Lambda_i} \circ {\mathrm{Tr}}_{\Lambda_i^\prime}=
{\mathrm{Tr}}_{(P^\prime\cap \Lambda_i)\cup \Lambda_i^\prime}= {\mathrm{Tr}}_{(P\cap\Lambda_i)^\prime}.$
Where we used $(P^\prime\cap \Lambda_i)\cup \Lambda_i^\prime= (P^\prime\cup \Lambda_i^\prime)\cap (\Lambda_i\cup \Lambda_i^\prime)=
 P^\prime\cup \Lambda_i^\prime=(P\cap \Lambda_i)^\prime=P_i^\prime.$ 
\vskip 0.2truecm
If in Eq.~(\ref{Phi}) one considers just a subset of partitions of $\Lambda$ one finds an upper-bound of $\Phi.$ 
One can define a monotonic family $\{\Phi^{(k)}\}_{k=1}^{|\Lambda|}$ of II measures by defining $\Phi^{(k)}$ as in Eq.~(\ref{Phi}) where the minimization
is performed over partitions ($\Lambda_1, \Lambda_1^\prime)$ in which $|\Lambda_1|\le k.$ Notice that
$i\in\Lambda$) and that $i\ge j\Rightarrow \Phi^{(j)}({\cal U})  \ge\Phi^{(i)}({\cal U})\ge\Phi^{(|\Lambda|)}=\Phi ,\, (i,j=1,\ldots,|\Lambda|,\forall{\cal U}).$ In our numerical experiments we found that often $\Phi^{(1)}=\Phi$. When this is the case the minimization in Eq.~(\ref{Phi}) requires to consider $O(|\Lambda|)$ partitions only.



\begin{thebibliography}{99}
\bibitem{time-reversal} In Fact, from Eq.~(\ref{repertoire}) one sees that causes and effects are exchanged by replacing $\cal U$ with its dual ${\cal U}^*$,
moreover $\varphi(M)$ is symmetric under this exchange ($\forall M$). Therefore, $C({\cal U}^*)=X C({\cal U}^*) X^\dagger$,
where $X:={\mathbf{1}}\otimes \sigma^x\otimes {\mathbf{1}}$. Since the distance $D$ in Eq.~(\ref{Phi}) is unitarily-invariant one gets $\Phi({\cal U}^*)=\Phi({\cal U})$.
\bibitem{group-action} Since $\rho^{(x)}_{(\otimes_i{\cal U}_i){\cal U} }(P|M)=(\otimes_i{\cal U}_i)(\rho^{(x)}_{{\cal U} }(P|M)),\,(\forall P,M)\Rightarrow\varphi_{(\otimes_i{\cal U}_i){\cal U} }(M)=
\varphi_{{\cal U} }(M)$ ($\|\bullet\|_1$-invariance under unitaries) one has that $C((\otimes_i{\cal U}_i){\cal U})=({\mathbf{1}}\otimes (\otimes_i{\cal U}_i))C({\cal U}),\,\forall{\cal U}$
i.e., conceptual structure operators transform covariantly, from Eq.~(\ref{CS-dist}), (\ref{Phi}), and using again  $\|\bullet\|_1$-invariance one obtains $\Phi((\otimes_i{\cal U}_i){\cal U})=\Phi({\cal U}).$
\bibitem{upper-bound}
By definition $\Phi({\cal U})\le D[ C({\cal U}), C({\cal U}_{\cal P})]$,  for all partitions $\cal P$.
From Eq.~(\ref{CS-dist}) one has $D[ C({\cal U}), C({\cal U}_{\cal P})]= \frac{1}{2}\| C({\cal U}) -C({\cal U}_{\cal P}) \|_1\le \frac{1}{2} (\| C({\cal U}) \|_1+\| C({\cal U}_{\cal P}) \|_1\le \frac{1}{2} \sum_{M\subset \Lambda} (\varphi_{\cal U}(M)+ \varphi_{{\cal U}_{\cal P}}(M) )\le \frac{1}{2}  (N_c({\cal U}) +N_c({\cal U}_{\cal P}) )\le N_c({\cal U})$. Here we have used the fact that the number of concepts $N_c({\cal U}_{\cal P})$ of the partitioned network is always not larger than the one $N_c({\cal U})$ of the un-partitioned one $\forall{\cal P}$. In the same way one proves $ \Phi({\cal U})\le  \sum_{M\subset \Lambda} \varphi_{\cal U}(M)$.
\bibitem{lower-bound} From the Definitions(\ref{CS-dist}), (\ref{Phi}) and the dis-integration Lemma in the Appendix  one has
$\Phi=D(C({\cal U}),C({\cal U}_{MIP}))=\frac{1}{4}\sum_{M,\,\alpha=e,c}^\prime\|\varphi_{\cal U}(M)\rho_{\cal U}^{(\alpha)}(M)-\varphi_{{\cal U}_{MIP}}(M)
 \rho_{{\cal U}_{MIP}}^{(\alpha)}(M)\|_1,$ where the sum is over those $M\subset\Lambda$ such that either $M$ or its core cause/effect are in ${\cal P}_{MIP}$
(as all the other $M$'s give contribution which  are the same for ${\cal U}$ and ${\cal U}_{MIP}$ and therefore cancel out.) In particular the sum is lower bounded
by restricting to the subset of $M\in\partial{\cal P}_{MIP}.$ Since for these dis-integrated mechanisms $\varphi_{{\cal U}_{MIP}}(M)=0$ 
and $\|\rho_{\cal U}^{(\alpha)}(M)\|_1=1$ one arrives at the first inequality Eq.~(\ref{Phi-lower-bound}). The second inequality is obvious.
\end{thebibliography}
\end{document}